\documentclass[Times,twocolumn,deluxetable]{aastex631}
\usepackage{graphicx}
\usepackage{multirow}
\usepackage{txfonts}
\usepackage{natbib}
\usepackage{url}
\usepackage{hyperref}
\usepackage{longtable}
\usepackage{tabularx}
\usepackage{lipsum}
\usepackage{array}
\usepackage{rotating}
\usepackage{changepage}

\submitjournal{AAS}

\shorttitle{X-ray Properties of GS 1354-64}
\shortauthors{K. Chatterjee et al.}

\begin{document}

\title{Mapping the Contracting Corona in GS 1354-64 with Insight-HXMT and XRISM: Evolving X-ray Lags and Quenched Reflection across a Complete State Transition}

\author[0000-0002-6252-3750]{Kaushik Chatterjee$^\dagger$}
\affiliation{South-Western Institue For Astronomy Research, Yunnan University, University Town, Chenggong, Kunming 650500, Yunnan, People's Republic of China}
\affiliation{Key Laboratory of Survey Science of Yunnan Province, Yunnan University, Kunming 650500, Yunnan, People's Republic of China}

\author[0009-0008-5223-2934]{Min Wei$^\ddagger$}
\affiliation{Department of Astronomy, School of Physics and Astronomy, Yunnan University, University Town, Chenggong, Kunming 650500, Yunnan, People's Republic of China}

\author[0009-0007-7562-7720]{Jie Liao$^\ddagger$}
\affiliation{School of Physics and Astronomy, Sun Yat-Sen University, Zhuhai 519082, Guangdong, People's Republic of China}

\author[0000-0002-4533-3170]{Biswaraj Palit$^\ddagger$}
\affiliation{Nicolaus Copernicus Astronomical Center, Polish Academy of Sciences, ul. Bartycka 18, 00-716, Warsaw, Poland}

\author[0000-0003-0793-6066]{Santanu Mondal}
\affiliation{Department of Physics and Electronics, Christ University, Hosur Main Road, Bengaluru, 560029, India}

\author[0000-0003-2845-1009]{Honghui Liu}
\affiliation{Institut f\"ur Astronomie und Astrophysik, Eberhard-Karls Universit\"at T\"ubingen, D-72076 T\"ubingen, Germany}

\author[0000-0002-7782-5719]{Chandra B. Singh}
\affiliation{Calle 134A No. 148A 21 San Pedro de Tibabuyes, Suba - Bogota, Colombia}

\author[0000-0001-7590-5099]{Sandeep K. Rout}
\affiliation{New York University Abu Dhabi, PO Box 129188, Abu Dhabi, UAE}
\affiliation{Center for Astrophysics and Space Science (CASS), New York University Abu Dhabi, PO Box 129188, Abu Dhabi, UAE}

\author[0000-0002-3960-5870]{Ashutosh Tripathi}
\affiliation{Xinjiang Astronomical Observatory, Chinese Academy of Sciences, 150 Science-1 Street, Urumqi, 830011, People's Republic of China}

\author[0000-0002-5900-9785]{Mayukh Pahari}
\affiliation{Department of Physics, Indian Institute of Technology Hyderabad, Hyderabad, Kandi, 502285 Sangareddy, India}

\author[0000-0001-8630-5435]{Guobao Zhang}
\affiliation{Yunnan Astronomical Observatories, Chinese Academy of Sciences, Kunming 650216, People's Republic of China}

\author[0000-0002-4705-2479]{Zejun Jiang}
\affiliation{Department of Astronomy, School of Physics and Astronomy, Yunnan University, University Town, Chenggong, Kunming 650500, Yunnan, People's Republic of China}

\author[0000-0001-7225-2475]{Brajesh Kumar}
\affiliation{South-Western Institue For Astronomy Research, Yunnan University, University Town, Chenggong, Kunming 650500, Yunnan, People's Republic of China}
\affiliation{Key Laboratory of Survey Science of Yunnan Province, Yunnan University, Kunming 650500, Yunnan, People's Republic of China}

\author[0000-0003-3901-8403]{Wei Wang}
\affiliation{Department of Astronomy, School of Physics and Technology, Wuhan University, Wuhan, People's Republic of China}

\author[0000-0002-5617-3117]{Hsiang-Kuang Chang}
\affiliation{Institute of Astronomy, National Tsing Hua University, Hsinchu 300044, Taiwan, Republic of China}

\author[0000-0003-1295-2909]{Xiaowei Liu}
\affiliation{South-Western Institue For Astronomy Research, Yunnan University, University Town, Chenggong, Kunming 650500, Yunnan, People's Republic of China}
\affiliation{Key Laboratory of Survey Science of Yunnan Province, Yunnan University, Kunming 650500, Yunnan, People's Republic of China}

\let\thefootnote\relax\footnotetext{$^\dagger$Corresponding Author (Email ID: \href{mailto:mails.kc.physics@gmail.com}{mails.kc.physics@gmail.com})}
\let\thefootnote\relax\footnotetext{$^\ddagger$These authors have contributed equally to this work}

\begin{abstract}

We present a comprehensive broadband spectro-temporal analysis of the dynamically confirmed Galactic black hole transient GS 1354-64 during its 2025--26 outburst, using near-simultaneous observations from \textit{Insight}-HXMT and \textit{XRISM}. The high spectral resolution of the \textit{XRISM}-Resolve microcalorimeter, coupled with the high-cadence broadband coverage of \textit{Insight}-HXMT, reveals a dramatic evolution in the inner accretion geometry. During the early hard-intermediate state, the spectrum was dominated by a hot Comptonizing corona accompanied by relativistically broadened reflection features. As the outburst progressed, the corona cooled efficiently, the photon index steepened ($\Gamma \gtrsim 2.5$), and the reflection signatures, including the broad Fe K$\alpha$ emission, were quenched, yielding a featureless, disk-dominated continuum. Concurrently, Type-C quasi-periodic oscillations (QPOs) evolved from $0.25$~Hz to $6.52$~Hz. Cross-spectral timing analysis revealed that the broadband X-ray time lags changed from pronounced positive (hard) delays ($\sim 160$~ms) at low frequencies to near zero as the source softened. However, a statistical analysis incorporating the measurement uncertainties shows that the broadband lags lack a significant monotonic correlation with the QPO frequency, photon index, or fractional rms. Instead, the scattered lag behavior suggests a competition between positive Comptonization delays and negative thermal reverberation. While the spectral softening and QPO acceleration provide strong evidence for a geometric contraction of the corona---whether interpreted via Lense--Thirring precession or the Two-Component Advective Flow (TCAF) paradigm---the broadband timing properties underscore the turbulent and inhomogeneous nature of the inner flow as the advancing Keplerian disk drives the source into the soft state.

\end{abstract}


\section{Introduction}
\label{sec:intro}

Accreting stellar-mass black holes in X-ray binaries (BHXBs) are laboratories for studying matter in strong gravity. Their X-ray spectra are generally described by two components: multi-temperature thermal emission from an optically thick, geometrically thin accretion disk \citep{Shakura1973}, and a power-law tail produced when soft disk photons are inverse-Compton scattered in a hot, optically thin electron plasma, the corona \citep{Sunyaev1980}.

During an outburst, transient BHXBs pass through distinct spectral states and trace a characteristic ``q''-shaped track in the hardness--intensity diagram (HID; \citealt{Remillard2006}). In the hard state (HS), the spectrum is dominated by Comptonization ($\Gamma \sim 1.4$--1.8) and the disk is thought to be truncated well outside the innermost stable circular orbit (ISCO). As the mass accretion rate rises, the disk moves inward and the source passes through the hard- and soft-intermediate states (HIMS and SIMS) into the soft state (SS), in which the disk extends to, or close to, the ISCO and dominates the X-ray emission while the coronal contribution is weak. During the decay, the source returns to the hard state at lower luminosity as viscous depletion and an inward-propagating cooling front drain the disk \citep{Tetarenko2020}.

The spectral evolution is closely coupled to the fast variability. The power density spectra (PDS) of BHXBs show broadband noise and low-frequency quasi-periodic oscillations (LFQPOs). Type-C QPOs, seen mainly in the HS and HIMS, are narrow features whose centroid frequency increases as the spectrum softens \citep{Casella2005}. A leading interpretation attributes them to Lense--Thirring precession, first proposed for test-particle orbits \citep{Stella1998} and later extended to a hot inner flow that precesses as a solid body inside a truncated disk \citep{Ingram2009}. In the Two-Component Advective Flow (TCAF) model \citep{Chakrabarti1995}, they instead arise from oscillations of the post-shock CENtrifugal pressure supported BOundary Layer (CENBOL), whose size, set by the interplay between the sub-Keplerian halo and the Keplerian disk, regulates both the spectral hardness and the QPO frequency \citep{Mondal2014, Mondal2015, Mondal2023}.

State transitions also change the outflows. Compact radio jets are present in the hard state and quenched in the soft state \citep{Fender2004}, while highly ionized disk winds, seen as narrow Fe\,\textsc{xxv} and Fe\,\textsc{xxvi} absorption lines, are detected mostly in soft states \citep{Neilsen2023}. The \textit{XRISM}/Resolve microcalorimeter \citep{Tashiro2021} can separate such narrow features from the broad relativistic reflection component in the Fe K band.

GS 1354-64 (BW Cir) is a dynamically confirmed BHXB with an orbital period of $\approx 2.54$~d and a mass function of $5.7 \pm 0.3\,M_\odot$ \citep{Casares2009}. Strong accretion-disk variability in quiescence prevented an inclination measurement, but the absence of X-ray eclipses gives a secure lower limit of $M_1 \geq 7.6 \pm 0.7\,M_\odot$ \citep{Casares2009}; \citet{Mondal2021} estimated $M_{\rm BH} \sim 7.0^{+0.5}_{-0.8}\,M_\odot$ from TCAF and JeTCAF modeling. The distance is uncertain: the properties of the donor star imply $d \gtrsim 25$~kpc \citep{Casares2009}, whereas \textit{Gaia} DR2 astrometry gives $0.7^{+3.4}_{-0.3}$~kpc \citep{Gandhi2019}.

The source was discovered by \textit{Ginga} in 1987, when its spectrum, a soft disk component with a power-law tail, was typical of the soft state \citep{Kitamoto1990}. The 1997 and 2015 outbursts, in contrast, remained in the hard state throughout \citep{Revnivtsev2000, Brocksopp2001, Koljonen2016, ElBatal2016}. During the 2015 outburst, \citet{Pahari2017} detected a $\sim$18~mHz QPO simultaneously in X-rays (\textit{Swift}/XRT) and in the optical (SALT), which they interpreted as evidence for optical cyclo-synchrotron emission from the hot accretion flow. The current outburst was detected in late 2025 \citep{Negoro2025, Rout2026}. Unlike in 1997 and 2015, the source began a hard-to-soft transition around MJD 61059 \citep{Negoro2026a}, reached the soft state after a secondary flux peak in mid-February 2026 \citep{Negoro2026b}, and returned to the hard state during the decay \citep{Adegoke2026b}. This is the first outburst since 1987 in which GS 1354-64 has reached the soft state, and the first in which its hard-to-soft transition has been followed with sensitive broadband timing and high-resolution X-ray spectroscopy \citep[see also][]{Adegoke2026a}.

In \citet{Chatterjee2026b} (hereafter Paper~I), we studied the early part of this outburst with a broadband multi-mission campaign. During the hard-state rise, narrow Type-C QPOs evolved from 0.08 to 1.73~Hz. The QPO frequency was anti-correlated with the inner disk radius ($R_{\rm in}$) and correlated with the photon index ($\Gamma$), consistent with an inward-moving disk that raises the precession frequency while cooling the corona. A highly ionized (Fe\,\textsc{xxvi}) equatorial disk wind was detected before the state transition. Relativistic reflection modeling indicated a near-maximally spinning black hole ($a_* > 0.99$) viewed at $i \approx 52^{\circ}$--$62^{\circ}$, with the inner disk at the ISCO. After the source entered a luminous, thermally dominated state, the QPOs moved to 4.7--6.2~Hz and weakened.

Here we extend Paper~I through the state transition and the subsequent evolution, using the high-cadence broadband monitoring of \textit{Insight}-HXMT and two \textit{XRISM}/Resolve observations, one at the onset of the transition and one about a month later. We follow the continuum and reflection properties, the QPOs, and the energy-dependent X-ray lags across the outburst. Section~\ref{sec:data} describes the observations and data reduction, Section~\ref{sec:analysis} the analysis methods, and Section~\ref{sec:results} the results, which we discuss in Section~\ref{sec:discussion} and summarize in Section~\ref{sec:summary}.

\section{Observations and Data Reduction}
\label{sec:data}

The 2025--2026 outburst of GS 1354-64 was extensively monitored by a multitude of X-ray observatories. To investigate the broadband spectral and temporal evolution of the source across its state transitions, we utilize data from \textit{XRISM} and \textit{Insight}-HXMT. The comprehensive observation logs for all instruments are summarized in Table \ref{tab:obs_log}.

\subsection{Data Reduction}

\subsubsection{XRISM}
\label{subsec:xrism}
Archival data for the Galactic black hole transient GS 1354-64 were obtained via the public \href{https://heasarc.gsfc.nasa.gov/cgi-bin/W3Browse/w3browse.pl}{HEASARC archive}. The data are reduced following the standard procedure described in the XRISM Data Analysis Guide\footnote{\url{https://heasarc.gsfc.nasa.gov/docs/xrism/analysis/}}. We use HEASOFT v6.36 together with CALDB version 20250915. Spectra and light curves are extracted with \texttt{xselect}.

For Resolve, we select only high-resolution primary events with \texttt{GRADE=0}, and exclude events from pixels 12 and 27. The extra-large (X) response matrix is generated using \texttt{rslmkrmf}. We then produce the exposure map with \texttt{xaexpmap}, and generate the Auxiliary Response File (ARF) using \texttt{xaarfgen}.

\subsubsection{Insight-HXMT}
\label{subsec:hxmt}

Raw Level-1 data for the outburst of GS 1354-64 were retrieved from the official archive and processed into Level-2 data products using the official \href{http://hxmten.ihep.ac.cn/software.jhtml}{HXMTDAS} software suite (version 2.07). The primary data reduction was executed via the \texttt{hpipeline} script, with Good Time Interval (GTI) filtering requiring an elevation angle $> 10^\circ$, a geomagnetic cutoff rigidity $> 8$~GeV, a pointing offset angle $< 0.04^\circ$, and an exclusion window of at least 600~s around passages through the South Atlantic Anomaly (SAA). To minimize background contamination, we selected events only from the small field-of-view (FOV) detectors of all three payloads, as recommended in the \href{http://hxmten.ihep.ac.cn/SoftDoc/501.jhtml}{HXMT Data Analysis Manual}. Instrument-specific spectra and their response matrices were extracted using \texttt{hespecgen}, \texttt{mespecgen}, and \texttt{lespecgen}, alongside \texttt{herspgen}, \texttt{merspgen}, and \texttt{lerspgen}, respectively. Light curves for the timing analysis were extracted via the \texttt{helcgen}, \texttt{melcgen}, and \texttt{lelcgen} tasks. Background models for both the temporal and spectral products were estimated with the \texttt{hebkgmap}, \texttt{mebkgmap}, and \texttt{lebkgmap} tools. To ensure the validity of $\chi^2$ statistics in our \texttt{XSPEC} modeling, the source spectra were rebinned with the FTOOLS \texttt{grppha} task to contain a minimum of 30 counts per bin. Standard systematic errors of 1\% and 2\% were added for the LE and ME instruments, while channel-dependent systematic errors were added for the HE instrument from the standard `\texttt{HEsyserror.dat}' file. For the extraction of PDS and the search for QPOs, the light curves were generated with a time resolution of $0.0078125$~s.

\section{Data Analysis}
\label{sec:analysis}

\subsection{Spectral Analysis}
\label{subsec:spectral_analysis}

\subsubsection{Broadband HXMT Data}
\label{subsubsec:broadband_spec-analysis}
\begin{figure}[ht!]
      \centering
      \includegraphics[width=\columnwidth]{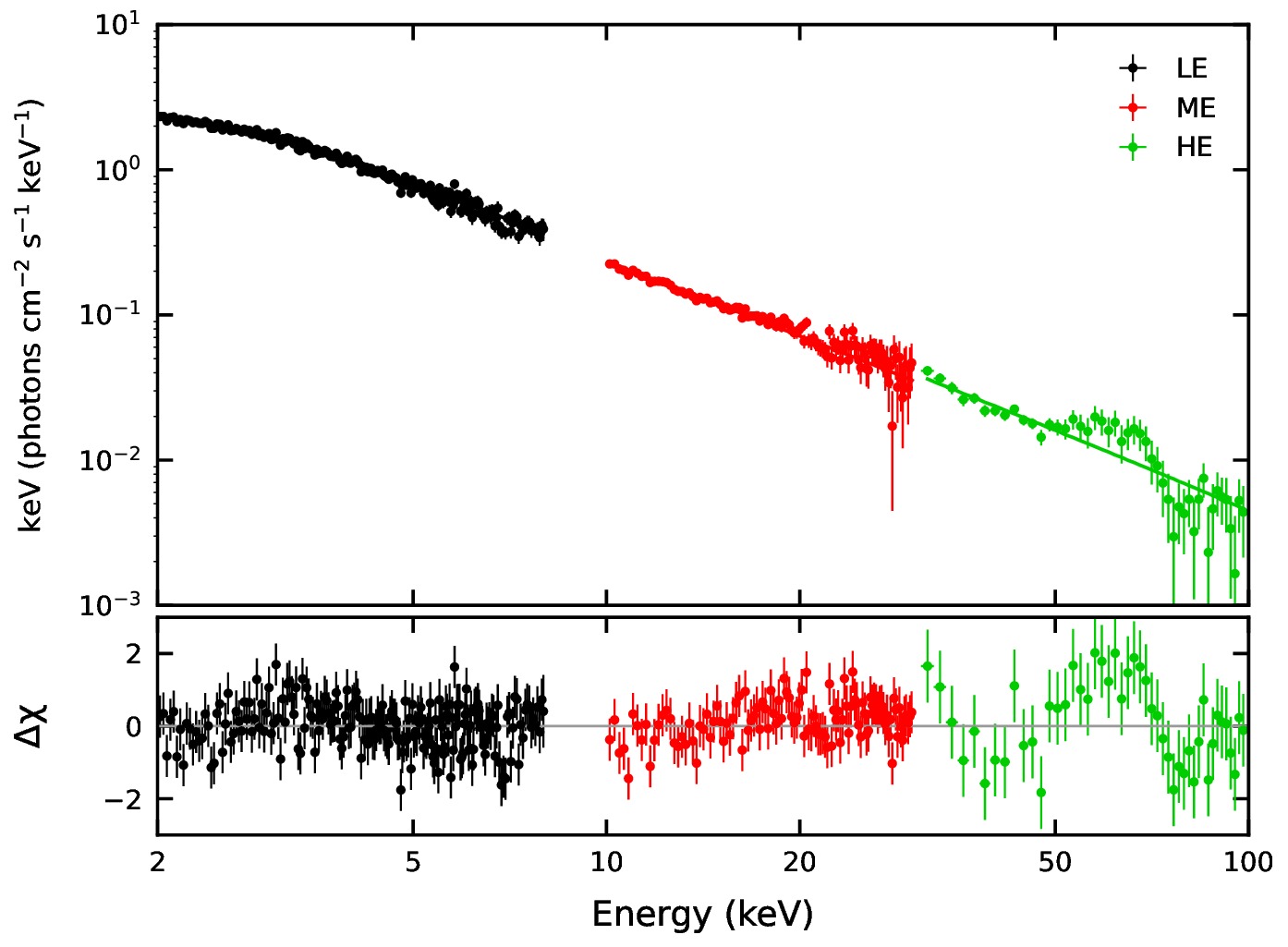}
      \caption{Representative broadband spectrum of \textit{Insight}-HXMT data (ExpID P081404800303) fitted with
      \texttt{constant*tbabs*cutoffpl}. The LE, ME, and HE spectra were fitted jointly over 2--8, 10--30, and 30--100~keV, respectively.
      The upper panel shows the unfolded spectrum and best-fit model, and the lower panel shows the residuals with the reduced $\chi^2$ is 0.92.}
      \label{fig:cutoffpl_spectrum}
  \end{figure}

We performed broadband spectral fitting of the \textit{Insight}-HXMT LE, ME, and HE data on an exposure-by-exposure basis using \texttt{XSPEC} v12.13.1 \citep{Arnaud1996}. For each epoch, the grouped spectra from the three instruments were fitted jointly. To minimize the influence of background systematics and focus on stable source channels, the fitting ranges were constrained to 2--8~keV for the LE, 10--30~keV for the ME, and 30--100~keV for the HE payloads. A cross-normalization constant was applied to account for inter-instrument calibration uncertainties; the LE constant was fixed at unity, while the ME and HE constants were permitted to vary freely.

The spectral modeling was conducted chronologically across the outburst to track the continuum evolution, rather than binning the observations into predefined, static spectral states. We first adopted simple phenomenological continuum models capable of adequately describing the emission at each epoch, and also modeled the spectra with the relativistic \texttt{relxill} model \citep{Garcia2014, Dauser2014}.

For the Compton-dominated spectra, we applied the composite model \texttt{constant*tbabs*relxill}, which can describe both the primary hard X-ray continuum and any associated broad curvature. In the baseline \texttt{relxill} fits, the black hole spin ($a_* = 0.998$), inclination ($i = 50^\circ$), break radius ($15\,r_g$), outer disk radius ($400\,r_g$), redshift ($z=0$), high-energy cutoff ($E_{\rm cut} = 300$~keV), emissivity index (\texttt{Index1}, with \texttt{Index2} tied to it), iron abundance ($A_{\rm Fe}=1$), and inner disk radius ($R_{\rm in}$) were fixed, while the neutral hydrogen column density ($N_{\rm H}$), coronal photon index ($\Gamma$), ionization parameter ($\log\xi$), reflection fraction ($R_{\rm refl}$), and the overall normalization were left free. The frozen values were chosen following previous \texttt{relxill} studies of black hole X-ray binaries \citep{Liu2026}.

For a small subset of spectra, the baseline model left soft residuals that required an additional thermal disk component. These spectra were therefore refitted with \texttt{constant*tbabs*(diskbb+relxill)}, in which the disk temperature and disk normalization were allowed to vary in addition to $N_{\rm H}$, $\Gamma$, $\log\xi$, $R_{\rm refl}$, and the overall \texttt{relxill} normalization.

For the thermally dominated epochs, the continuum was best described by a two-component model, \texttt{constant*tbabs*(diskbb + powerlaw)}, with $N_{\rm H}$, the inner disk temperature ($kT_{\rm in}$), the disk normalization ($N_{\rm diskbb}$), the power-law photon index ($\Gamma$), and the power-law normalization free. We tested alternative continuum combinations for these epochs (e.g., \texttt{diskbb + cutoffpl}), but these yielded only marginal statistical improvements. Consequently, the \texttt{relxill} parameters should be interpreted primarily as a phenomenological mapping of the hard-continuum curvature, while the \texttt{diskbb + powerlaw} model provides a more appropriate description of the continuum during the thermally dominated epochs. The derived parameters form the basis of the spectral evolution discussed in Section~\ref{subsubsec:broadband_properties}.

\subsubsection{High-Resolution XRISM Data}
\label{subsubsec:high-res_spec-analysis}
\begin{figure*}[ht!]
    \centering
    \includegraphics[width=\textwidth,height=6.5cm]{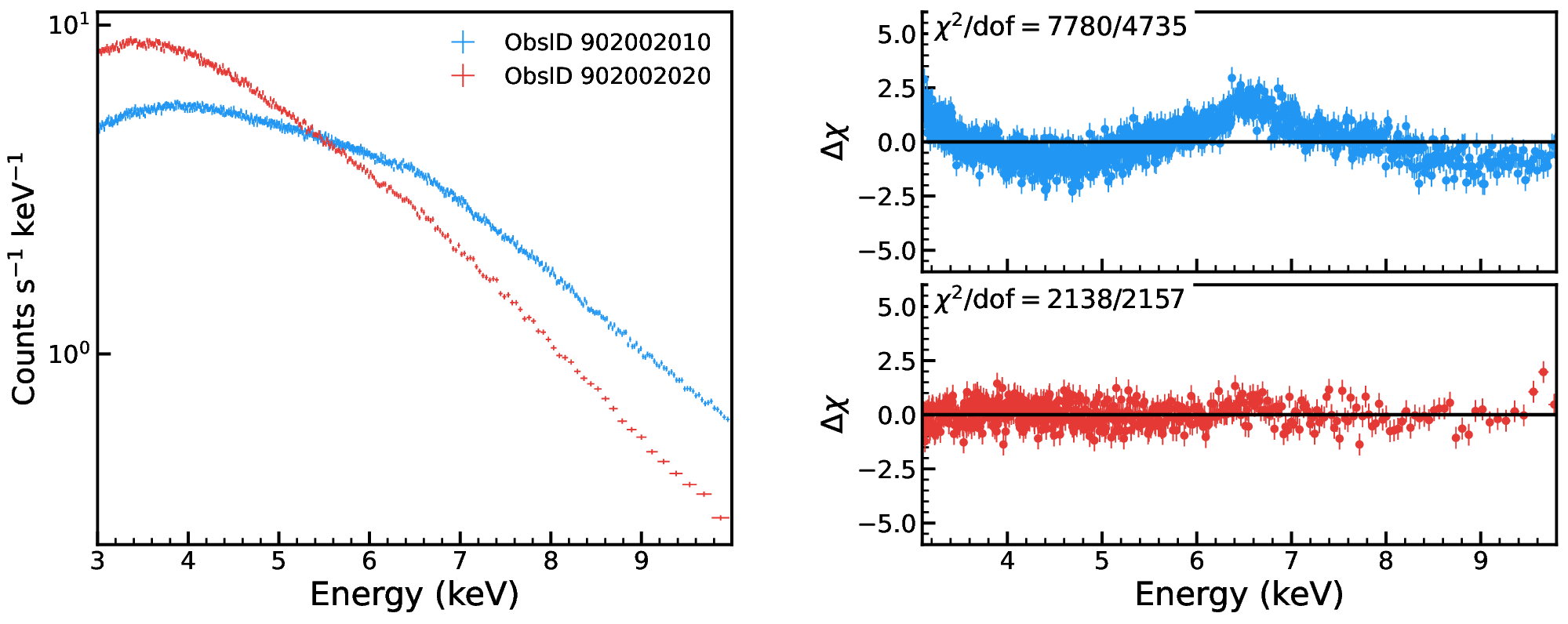}
    \caption{{\it Left}: The 3--10 keV {\it XRISM}-Resolve counts spectra corresponding to the two observations as listed in Tab.~\ref{tab:obs_log}. The observation ID 902002010 belongs to the hard state while the other observation ID 902002020 belongs to the soft state. {\it Right}: Residuals from the best-fitted \texttt{tbabs*cutoffpl} model to the 3 -- 10 keV Resolve spectra. The data in the plots are rebinned for visual clarity.}
    \label{fig:xrism_comparison}
\end{figure*}

We examined the two \textit{XRISM}-Resolve spectra obtained during the observational windows shown in Figure~\ref{fig:lightcurve} and listed in Table~\ref{tab:obs_log}, restricting the energy range to 3--10 keV. We further rebinned the data to a minimum of 1000 counts per bin using the \texttt{grppha} command, and used \texttt{XSPEC} v12.12.1 for the spectral analysis. The spectra of both epochs are shown in the left panel of Figure~\ref{fig:xrism_comparison}, where we observe a clear spectral pivoting between the two epochs. As an initial step, we fitted both spectra with a simple absorbed power-law model (\texttt{tbabs*powerlaw}) to identify prominent spectral features, particularly the broadened Fe K emission and disk reflection signatures. In both cases, we fixed the Galactic line-of-sight absorption column density at $7 \times 10^{21}$ cm$^{-2}$. This returned statistically unacceptable fits for both observations, leaving clear residuals around 6--7 keV as well as above 8 keV. Replacing the power-law continuum with the \texttt{cutoffpl} model improved the fit statistics only marginally. For the first spectrum (shown in blue in the top right panel of Figure~\ref{fig:xrism_comparison}), strong residuals persisted, suggestive of a broad Fe K line, consistent with recent work by \citet{Liu2026} and \citet{Chatterjee2026b}. In contrast, the second observation (shown in red in the bottom right panel of Figure~\ref{fig:xrism_comparison}) is adequately described by this simple phenomenological model. We did not detect any narrow Fe K emission or characteristic Fe K absorption features in either spectrum.

The first \textit{XRISM} observation was obtained simultaneously with \textit{NuSTAR}, extending the broadband spectral coverage up to 70 keV. Both Paper~I and \citet{Liu2026} performed joint spectral modeling, which revealed a rapidly spinning black hole together with a reflection-dominated spectrum. We adopted the best-fitting model from Paper~I, \texttt{const*tbabs*relxill}, and tested its applicability to the second \textit{XRISM} observation. To extend the spectral coverage beyond the \textit{XRISM} bandpass for the second epoch, we used quasi-simultaneous observations from the \textit{Insight}-HXMT low-, medium- and high-energy detectors (ObsID: P081404802603-20260219-01-01). We then fitted this model jointly and compared it with an alternative, simpler model, \texttt{const*tbabs*(diskbb+cutoffpl)}. The results of the spectral fitting and model comparison are described in Section~\ref{subsubsec:high-res_properties}.

\subsection{Timing Analysis}
\label{subsec:temporal_analysis}

\subsubsection{QPO Analysis}
\label{subsubsec:qpo-analysis}
\begin{figure*}[ht!]
    \centering
    \includegraphics[width=\textwidth,height=6.5cm]{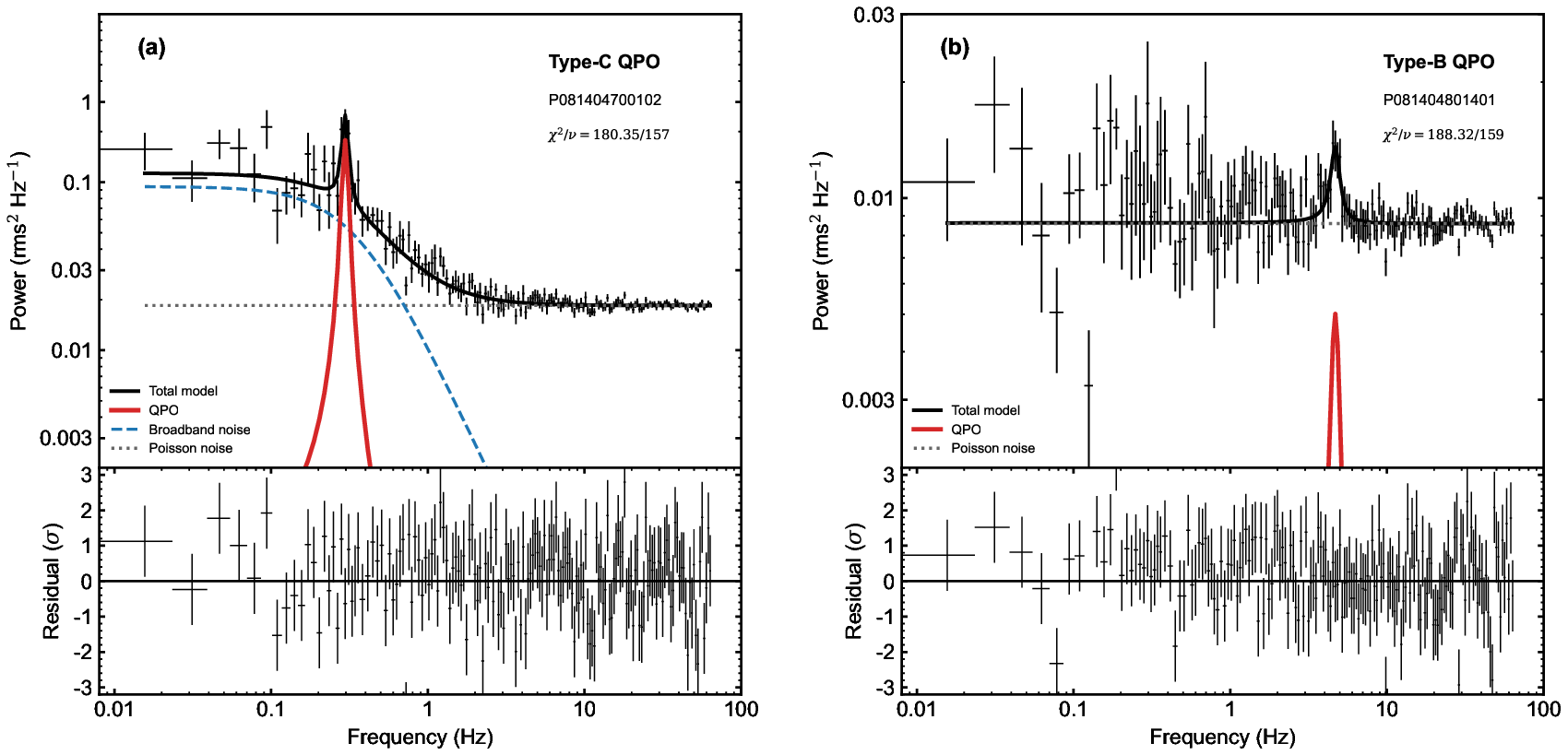}
    \caption{Power density spectra of GS\,1354-64 in the 2--10~keV band (i.e., the LE instrument). (a) Type-C QPO (ExpID P081404700102). (b) Type-B QPO (ExpID P081404801401). The bottom panels show the fit residuals. The best-fit models (black solid lines) consist of Lorentzian components for the QPO (red solid lines) and broadband noise (blue dashed lines, where applicable), plus constant/powerlaw components (dotted grey lines) for Poisson noise.}
    \label{fig:pds_comparison}
\end{figure*}

To investigate the fast temporal variability and extract the properties of the LFQPOs, we performed a timing analysis using the high-cadence data from the \textit{Insight}-HXMT LE, ME, and HE payloads. Source light curves were used to Nyquist-sample frequencies up to 64 Hz, sufficiently covering the dynamic range of typical BHXB fast variability. For each continuous GTI, the PDS were computed from the source light curves using \texttt{powspec} (HEAsoft) with \texttt{nbint=$8192$} and \texttt{nintfm=INDEF}, which divides the light curve into multiple statistically independent segments, computes a power spectrum for each segment, and averages them to reduce the statistical noise.

The resulting PDS were normalized according to the fractional root-mean-square (RMS) squared scheme \citep{Belloni1990, Miyamoto1991}, in which the integrated power over a frequency interval directly yields the squared fractional RMS amplitude of the source variability. The PDS were then geometrically rebinned in logarithmic frequency space ($\Delta f/f \approx 0.03$) to enhance the signal-to-noise ratio at higher frequencies.

To quantify the centroid frequencies ($\nu_{\rm QPO}$), widths, and RMS amplitudes of the timing features, the PDS were converted into a format compatible with standard spectral fitting tools and fitted in \texttt{XSPEC} (v12.13.1) \citep{Arnaud1996} using $\chi^2$ minimization. The broadband noise continuum was modeled with a combination of broad, zero-centered Lorentzian functions, and the narrow QPO fundamentals and their harmonics were fitted with Lorentzian profiles \citep{Belloni2002}. The Poisson noise level was modeled as a power-law component.

For a Lorentzian component defined by its centroid frequency ($\nu_0$), Full Width at Half Maximum (FWHM, $w$), and normalization ($K$), the coherence of the oscillation was parameterized by its Quality Factor, $Q = \nu_0 / w$. The fractional RMS amplitude of a Lorentzian component was calculated as $RMS = \sqrt{K}\times(S+B)/S$, where $S$ and $B$ are the source and background count rates, respectively, with the associated $1\sigma$ error bounds derived using the \texttt{error} command within \texttt{XSPEC}. Representative best-fit models are shown in Figure \ref{fig:pds_comparison}, and the results are discussed in Section~\ref{subsubsec:qpo_properties}.

\subsubsection{Lag Analysis}
\label{subsubsec:lag-analysis}

To probe the geometrical configuration and spatial extent of the inner accretion flow, we performed a cross-spectral timing analysis to compute both energy-resolved and broadband phase and time lags. High-resolution light curves with the same time bin size were extracted using the \texttt{hxmtpipeline}. To map the lag-energy spectra, we generated light curves in 18 narrow energy bins spanning the three payloads: 2--4, 4--6, 6--8, and 8--10~keV for LE; 10--15, 15--20, 20--25, and 25--30~keV for ME; and 30--35, 35--40, 40--45, 45--50, 50--60, 60--70, 70--80, 80--90, 90--100, and 100--120~keV for HE. To trace the temporal evolution over the outburst, we also extracted broadband light curves for the LE (4--10~keV), ME (10--30~keV), and HE (27--250~keV) bands. To avoid correlated Poisson noise, we used mutually exclusive (non-overlapping) energy bins. The soft, disk-dominated 2--4~keV LE band was used as the reference light curve for all cross-spectra.

We used the \texttt{Stingray} Python package \citep{Huppenkothen2019} to generate the cross-spectra. The subject and reference light curves were divided into identical, continuous segments of 44.5~s to calculate the averaged complex cross-spectrum ($C = S^* \times R$, where $S$ and $R$ are the Fourier transforms of the subject and reference bands, respectively; see \citealt{Nowak1999} and \citealt{Uttley2014} for formal definitions). This segment length gives a frequency resolution of $\sim 0.022$~Hz, sufficient to resolve the narrow QPO profiles, while providing enough photons per segment in the photon-starved HE band (27--250~keV) for coherent lag measurements. The time lag ($\tau$) was derived from the phase lag ($\Delta\phi = \arg[C]$) evaluated at the centroid frequency of the fundamental QPO, such that $\tau = \Delta\phi / (2\pi \nu_{\rm QPO})$. By this convention, a positive lag indicates that the hard (subject) photons are delayed relative to the soft (reference) photons. The results are discussed in Section~\ref{subsubsec:lag_results}.

\section{Results}
\label{sec:results}

\subsection{Evolution of the Outburst}
\label{subsec:outburst_evolution}

\begin{figure}[ht!]
    \centering
    \includegraphics[width=\columnwidth,height=11cm]{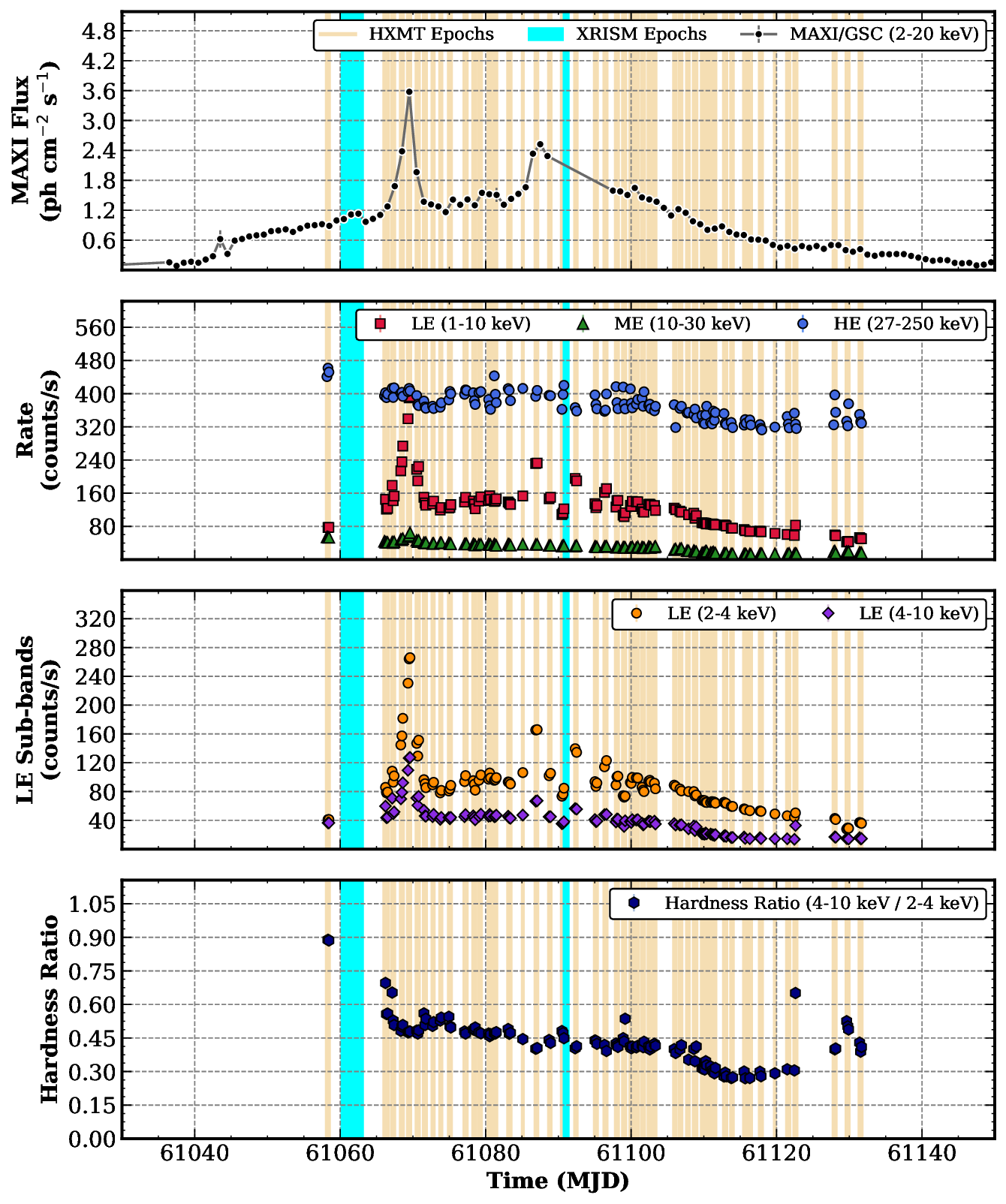}
    \caption{Broadband temporal evolution of GS 1354-64 during the 2025--2026 outburst. \textit{Top panel:} The overarching outburst envelope observed by MAXI/GSC (2--20 keV). \textit{Second panel:} Evolution of the \textit{Insight}-HXMT LE (1--10 keV), ME (10--30 keV), and HE (27--250 keV) count rates. \textit{Third panel:} Evolution of the \textit{Insight}-HXMT LE sub-bands (2--4 keV and 4--10 keV). \textit{Bottom panel:} The Hardness Ratio (4--10 keV / 2--4 keV). The light wheat-colored patches mark the epochs of the \textit{Insight}-HXMT monitoring, while the cyan patches denote the specific \textit{XRISM} observation windows.}
    \label{fig:lightcurve}
\end{figure}

\begin{figure}[ht!]
    \centering
    \includegraphics[width=\columnwidth]{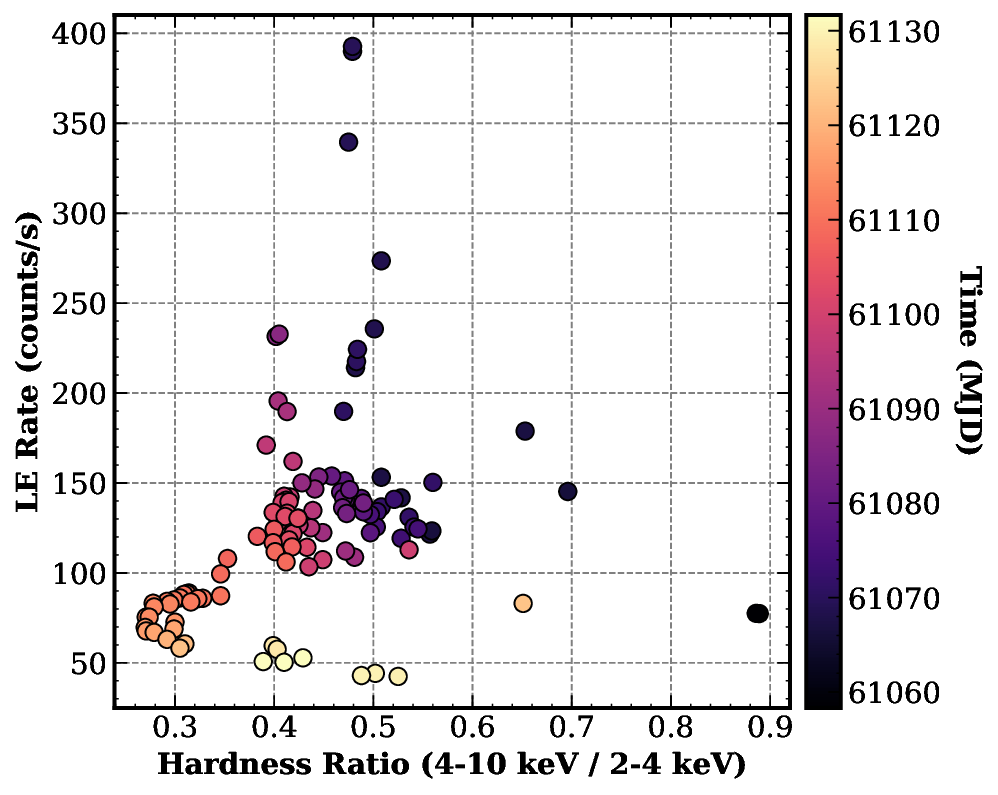}
    \caption{The Hardness-Intensity Diagram (HID) for GS 1354-64. The LE (1--10 keV) count rate is plotted against the Hardness Ratio (4--10 keV / 2--4 keV). The chronological color gradient (magma) tracks the source from its hard state rise into the soft state.}
    \label{fig:hid}
\end{figure}

To map the evolution of the 2025--2026 outburst of GS 1354-64, we generated broadband light curves using the \textit{Insight}-HXMT LE (1--10 keV), ME (10--30 keV), and HE (27--250 keV) payloads. Figure \ref{fig:lightcurve} tracks the multi-band flux evolution alongside the Hardness Ratio (HR), defined as the ratio of the LE sub-band count rates (4--10 keV / 2--4 keV). The overall 2--20 keV outburst envelope is captured by MAXI/GSC (Figure \ref{fig:lightcurve}, top panel). A detailed analysis of the MAXI/GSC light curve and of the initial state transitions is presented in Paper~I; here, we focus on the evolutionary phases covered by \textit{Insight}-HXMT and \textit{XRISM}.

The \textit{Insight}-HXMT monitoring (wheat-colored region) commenced in a bright, hard state (MJD $\sim 61058$). The high count rates in the HE band are due to its large effective area. At the onset of the monitoring, the 2--4 keV count rates were nearly equal to the 4--10 keV count rates in the LE band, but as the outburst progressed, the softer band started to dominate. Around the first \textit{XRISM} observation (first cyan patch, MJD 61060.15--61063.14), the soft X-ray (LE) flux rose sharply, driving a rapid initial drop in the hardness ratio. Over the subsequent weeks, encompassing the second \textit{XRISM} epoch (second cyan patch, MJD 61090.72--61091.42), the system exhibited a sustained, gradual softening. The hard X-ray flux steadily decayed, while the HR declined, reaching a minimum of $HR \approx 0.27$ near MJD 61115. This behavior is the classic signature of the inner accretion disk migrating inward and dominating the emission, cooling the coronal plasma through inverse-Compton scattering.

To classify the spectral states during these epochs, we plot the Hardness-Intensity Diagram (HID) in Figure \ref{fig:hid}. Using the LE rate as a proxy for intensity against the HR, the source traces a clear evolutionary path. Originating in the hard state (bottom right, darker markers), the source rapidly brightens as it begins to soften, and then continues to migrate leftward into the soft state as the intensity decays. This track confirms that the targeted \textit{Insight}-HXMT and \textit{XRISM} exposures captured the source departing the hard state and entering a thermal-dominated soft state.

\subsection{Spectral Properties}
\label{subsec:spectral_results}

\subsubsection{Broadband Properties}
\label{subsubsec:broadband_properties}
\begin{figure}[ht!]
    \centering
    \includegraphics[width=\columnwidth, height=18cm]{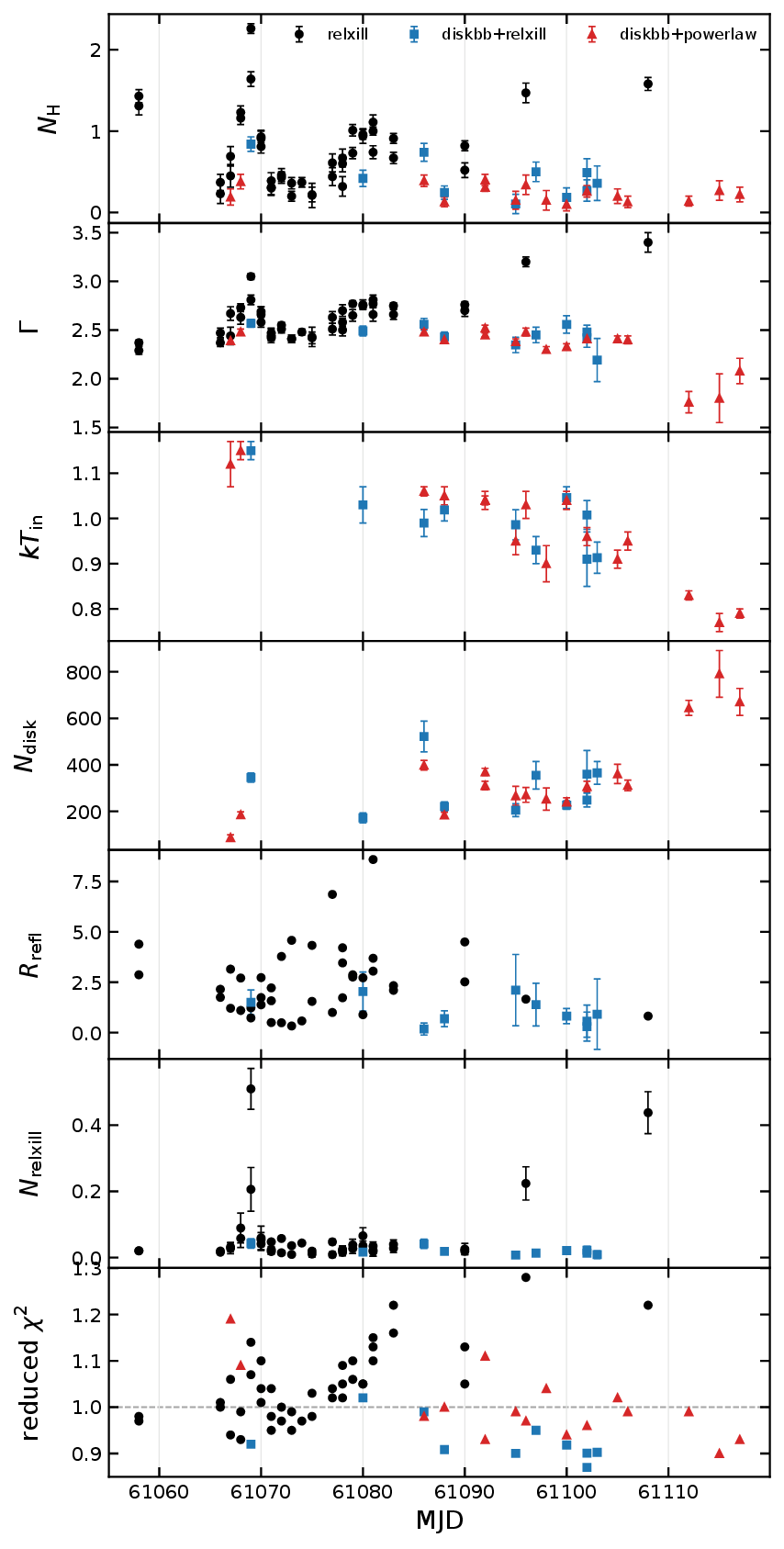}
    \caption{Temporal evolution of the main spectral parameters from the accepted \textit{Insight}-HXMT fits as a function of MJD. Symbols distinguish the adopted continuum models: \texttt{relxill},
    \texttt{diskbb+relxill}, and \texttt{diskbb+powerlaw}.}
    \label{fig:spectral_params_mjd}
\end{figure}

The evolution of the broadband spectral parameters derived from the \textit{Insight}-HXMT monitoring is summarized in Figure~\ref{fig:spectral_params_mjd} and Table~\ref{tab:spec_model_summary}.

The HXMT data in our analysis period begin when the outburst was already in the hard-intermediate state. In this initial hard-intermediate phase, the continuum was relatively flat, with a photon index of $\Gamma \simeq$ 2.4--2.5 in two observations on MJD 61066. The most rapid phase of spectral evolution occurred shortly thereafter (around MJD $\sim61070$), when the photon index steepened on a timescale of only a few days, increasing to $\sim$2.6 by MJD 61069 and peaking at $\Gamma \sim 3$ shortly after. This rapid transition indicates that the source quickly departed the hard branch toward a softer spectral configuration, corroborating the softening inferred from the HID evolution (Section~\ref{subsec:outburst_evolution}). We also observe a relatively broad variation in the fitted equivalent hydrogen column density ($N_{\rm H} \sim 0.1$--$2.3 \times 10^{22}$~cm$^{-2}$) across the transition; the potential physical and methodological drivers of this variation are addressed in Section~\ref{subsec:disc_spectral}.

Notably, we do not detect strong reflection signatures within the \textit{Insight}-HXMT spectra across these epochs. Specifically, the data lack a prominent, relativistically broadened Fe K$\alpha$ emission line or a clearly isolated high-energy Compton hump that would necessitate a reflection-dominated interpretation. One plausible explanation is that the rapidity of the hard-to-soft transition, unfolding over only $\sim$2--3 days, limited our observational cadence during the phase in which transient reflection features may have peaked in intensity.

We interpret the spectral residuals occasionally observed in the 50--80 keV band primarily as an instrumental feature rather than a physical Compton-reflection hump. As detailed in the \textit{Insight}-HXMT background calibration study by \citet{Liao2020}, the HE background is dominated by delayed radioactive activation induced by SAA passages, which produces a prominent, complex emission feature centered near 65~keV. Because the continuum flux of GS 1354-64 declines steeply at these energies, the spectrum is sensitive to small systematic uncertainties in the HE background subtraction, and residuals of this background line appear as a localized `hump' that does not require extreme reflection parameters.

The source, however, did not remain in a persistently soft spectral state after this brief softening excursion. Around MJD 61071, the photon index reversed its trend and decreased to $\Gamma \simeq 2.4$--2.5, a range that was maintained during MJD 61086--61106 in the acceptable \texttt{diskbb+powerlaw} fits. During this interval, the thermal component was clearly detected, with $kT_{\rm in} \simeq 0.90$--1.06~keV and $N_{\rm diskbb}$ values of a few hundred, indicating a substantial disk contribution to the soft X-ray emission. The coexistence of a prominent disk component and a still relatively steep power-law tail suggests that GS~1354$-$64 occupied a disk-dominated soft-intermediate state, rather than either a canonical hard state or a purely thermal-dominant soft state.

This disk-dominated phase was not sustained throughout the later evolution. As the outburst progressed, the inner disk temperature gradually decreased from $\sim1.04$~keV at MJD 61092 to 0.83~keV at MJD 61112, 0.77~keV at MJD 61115, and 0.65--0.69~keV by MJD 61121--61122. At the same time, the disk normalization increased to $N_{\rm diskbb}\sim(1.2$--$2.0)\times10^3$, consistent with a cooler and apparently larger disk-emitting region. Over the same late interval, the power-law continuum hardened, with $\Gamma$ decreasing from $\sim2.5$ to $\simeq1.7$--1.8. These trends indicate that the source evolved away from the disk-dominated soft-intermediate phase and returned toward a harder intermediate configuration. This is consistent with the timing behavior detailed in Section~\ref{subsec:temporal_results}: although the QPO frequency increased during the transition, the persistent, strong energy dependence of the fractional RMS variability shows that the inner, variable hard component remained dynamically dominant even when the disk emission became prominent.

\subsubsection{High-Resolution Properties}
\label{subsubsec:high-res_properties}

\begin{figure*}
    \centering
    \includegraphics[width=\textwidth,height=6.5cm]{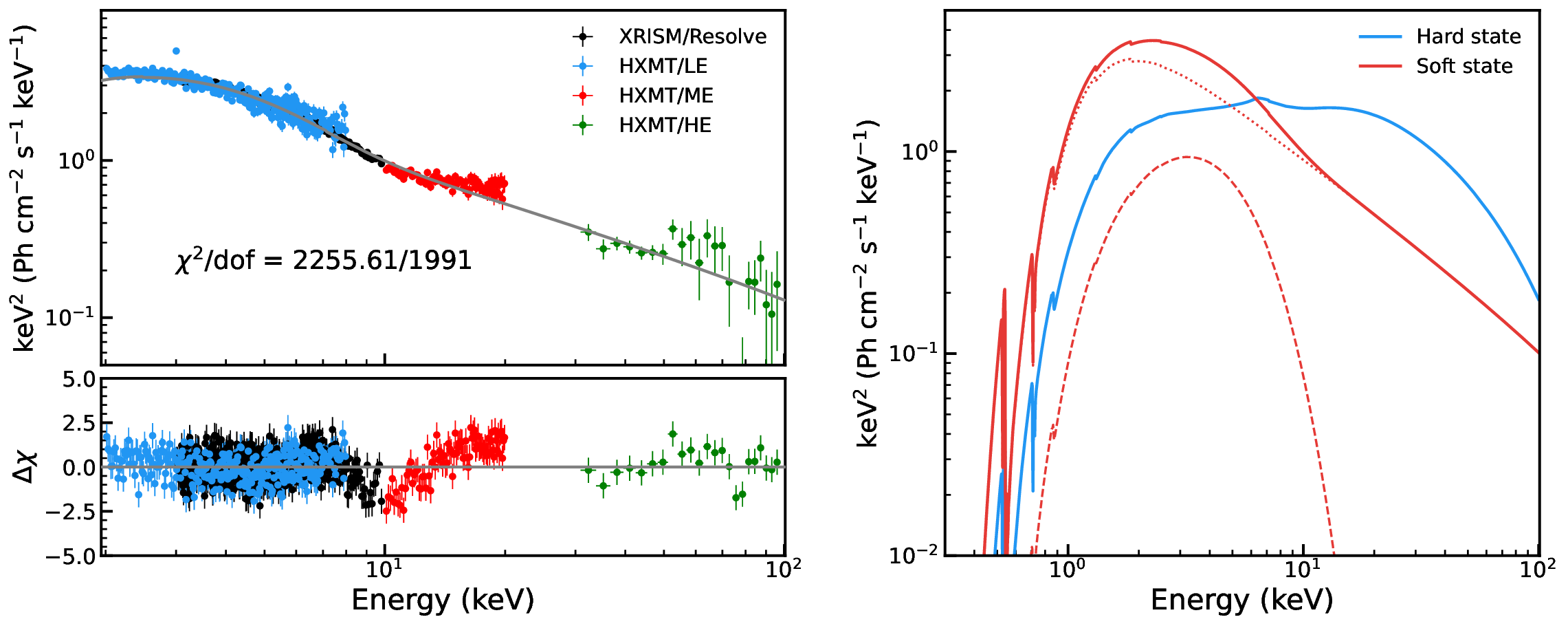}
    \caption{{\it Left}: Joint spectral modelling of {\it XRISM}-Resolve and {\it Insight}-HXMT fitted with \texttt{constant*tbabs*(diskbb+cutoffpl)} model. The {\it XRISM}, LE, ME, and HE spectra were fitted over 3--10 keV, 2--8 keV, 10--20 keV, and 30--100 keV, respectively. The data are re-binned further for visual clarity, and the best-fit model is shown as a gray curve. {\it Right}: A comparison of the best-fitting spectral models to both epochs, showing the spectral evolution in the source across the outburst episode.}
    \label{fig:model_compare}
\end{figure*}

A comparison of the two \textit{XRISM}-Resolve spectra in Figure~\ref{fig:xrism_comparison} indicates that, in the second epoch, the source exhibited enhanced emission at 3--5 keV together with a steeper high-energy tail. This spectral evolution is also evident from the time-resolved spectral analysis of the \textit{Insight}-HXMT observations, in which the disk-corona parameters undergo a systematic evolution (Figure~\ref{fig:spectral_params_mjd}). Our analysis of the high-resolution Resolve spectra supports this picture and demonstrates a clear transition from a reflection-dominated spectrum to a featureless continuum described by thermal disk emission and a power-law component: the broad Fe K emission feature detected in the first epoch is undetected in the second epoch.

We tested the best-fitting \texttt{const*tbabs*relxill} model of the first \textit{XRISM}+\textit{NuSTAR} spectrum from Paper~I (blue curve in the right panel of Figure~\ref{fig:model_compare}) on the second \textit{XRISM}+\textit{Insight}-HXMT observation. Although the reduced $\chi^2$ is $\sim 1$, both $R_{\rm refl}$ and the \texttt{relxill} normalization remained consistent with zero, indicating the absence of a statistically significant reflection component. In comparison, the alternative model \texttt{const*tbabs*(diskbb+cutoffpl)} provided an acceptable fit to the data, as presented in the left panel of Figure~\ref{fig:model_compare}. The best-fitting cross-calibration constants relative to \textit{XRISM}/Resolve were $0.66 \pm 0.01$, $1.23 \pm 0.01$, and $2.42 \pm 0.21$ for \textit{Insight}-HXMT/LE, ME, and HE, respectively. The $kT_{\rm in}$, $\Gamma$ and $N_{\rm diskbb}$ were constrained at $1.21 \pm 0.02$ keV, $2.87 \pm 0.02$ and $70.7 \pm 5.4$, respectively, consistent with the trends from the \textit{Insight}-HXMT observations described in Section~\ref{subsubsec:broadband_properties}. The comparison between the best-fitting models of the two epochs in the right panel of Figure~\ref{fig:model_compare} shows that, within $\sim 30$ days, the source evolved from a hard state (blue curve) to a soft, disk-dominated state (red curve). Between the two \textit{XRISM} observations, the 3--70 keV source flux increased by a factor of $\sim 2$ while $\Gamma$ steepened from 1.91 to 2.77. In the hard state, both Paper~I and \citet{Liu2026} measured a well-constrained high-energy cutoff of $E_{\rm cut}\approx40$ keV, whereas $E_{\rm cut}$ remained unconstrained during the soft-state observation and was therefore fixed at 500 keV. Together, these results indicate substantial changes in the properties of the hard X-ray emitting corona as the source evolved from the hard to the soft state.

\subsection{Temporal Properties}
\label{subsec:temporal_results}

\subsubsection{QPO Properties}
\label{subsubsec:qpo_properties}

\begin{figure}[ht!]
    \centering
    \includegraphics[width=\columnwidth,height=8.5cm]{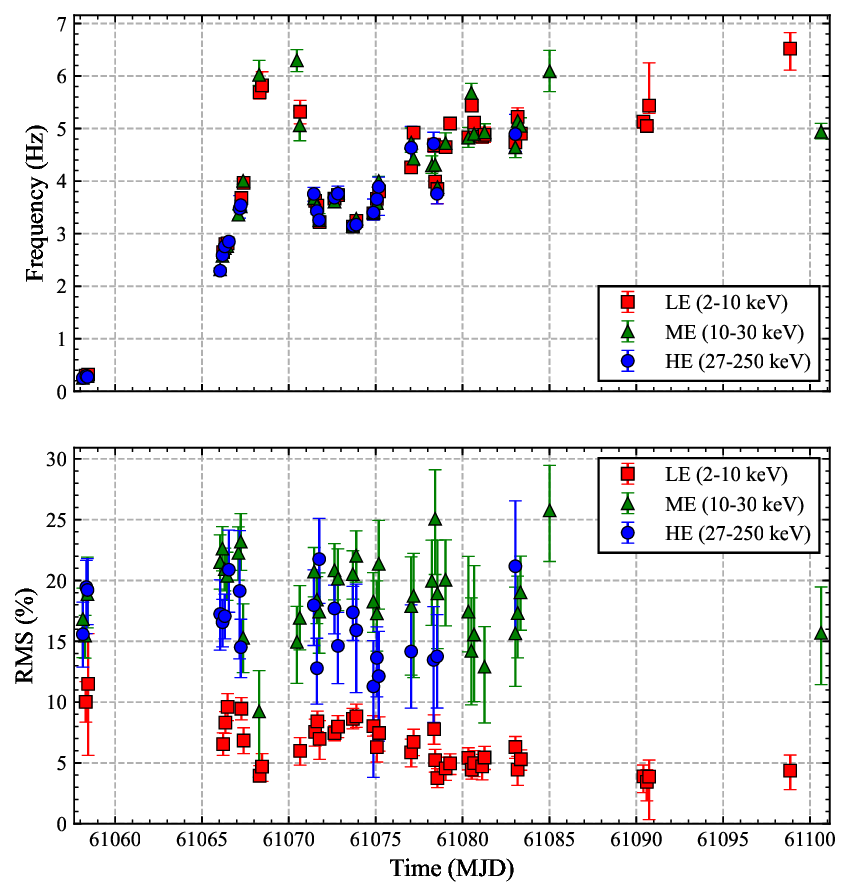}
    \caption{QPO frequency (top panel) and fractional RMS of the QPO component (bottom panel) evolution of GS 1354-64. The LE (2--10 keV), ME (10--30 keV), and HE (27--250 keV) bands are shown with red squares, green triangles, and blue circles, respectively. Error bars represent $1\sigma$ uncertainties.}
    \label{fig:freq_rms}
\end{figure}

We performed a timing analysis on the \textit{Insight}-HXMT exposures spanning MJD 61058--61132, covering the full hard-to-soft state transition and the subsequent decay (see Figure~\ref{fig:hid} for state classification). In total, 127 LE, 149 ME, and 147 HE individual exposures were obtained over this period. Requiring a Lorentzian component with a significance $\ge 3\sigma$ above the Poisson noise in the PDS fit, we detected QPOs in 39 LE, 38 ME, and 23 HE exposures within the narrower window MJD 61058--61101. The remaining exposures showed only noise-level variability in the respective bands and were excluded. The full set of best-fit QPO parameters is listed in Table~\ref{tab:qpo_params_final}.

The QPO frequency evolves with time across all bands (Figure~\ref{fig:freq_rms}, top panel). In the LE band, the QPO starts at $\nu_{\rm QPO} \approx 0.30$ Hz around MJD $\sim 61058$, rises to $\approx 3.68$ Hz by MJD $\sim 61067$ and $\approx 5.69$ Hz by MJD $\sim 61068$, and ultimately reaches a maximum of $\approx 6.52$ Hz near MJD $\sim 61098$. During the hard-to-intermediate transition (MJD $\sim 61066$--$61070$), the frequency increases rapidly, and the evolution then becomes more gradual during the soft intermediate state (MJD $\sim 61070$--$61084$). The centroid frequencies measured simultaneously in the ME and HE bands are consistent with the LE values within the $1\sigma$ uncertainties. The full frequency ranges are $0.30$--$6.52$ Hz (LE), $0.26$--$6.30$ Hz (ME), and $0.25$--$4.89$ Hz (HE). In the HE band, QPOs are detected in only 23 exposures; non-detections occur predominantly at $\nu_{\rm QPO} \gtrsim 5$ Hz, while the corresponding LE and ME observations show clear QPOs.

The fractional RMS amplitude varies with both frequency and energy band (Figure~\ref{fig:freq_rms}, bottom panel). In the LE band, the RMS ranges from $\sim 3.4\%$ to $\sim 11.5\%$; it is systematically higher in the ME band ($\sim 9.2\%$ to $\sim 25.8\%$) and spans $\sim 11.3\%$ to $\sim 21.8\%$ in the HE band. In the LE band, the RMS is highest at low QPO frequencies ($\nu \lesssim 1$ Hz) and decreases at higher frequencies ($\nu \gtrsim 5$ Hz). The amplification of the RMS at higher energies indicates that the physical mechanism driving the QPO is tied to the hot Comptonizing corona rather than to the thermally stable accretion disk \citep[see also][for similar energy-dependent QPO behaviors observed by \textit{Insight}-HXMT]{Wei2025}.

The quality factor $Q = \nu_{\rm QPO}/{\rm FWHM}$ ranges from $\sim 2.0$ to $\sim 22$ in the LE band, from $\sim 2.5$ to $\sim 26$ in the ME band, and from $\sim 3.3$ to $\sim 19$ in the HE band.

Three exposures (P081404801103, P081404801302, and P081404801401; MJD $\sim 61075$--$61078$) are identified as Type-B QPO candidates based on their occurrence within the SIMS phase (MJD $\sim 61070$--$61084$). Their LE QPO centroid frequencies span $\nu_{\rm QPO} \approx 3.8$--$4.9$ Hz, with LE RMS amplitudes of $\approx 6.7\%$--$7.8\%$ and $Q$ values of $\approx 6.6$--$15.4$ (see Figure~\ref{fig:pds_comparison} for a representative PDS). These values are consistent with the canonical properties of Type-B QPOs~\citep{Casella2005}. However, the boundaries between Type-B and Type-C QPOs in this source are not sharply defined: for instance, P081404801301 (MJD $\sim 61077$, adjacent to two of our Type-B candidates) shows an LE QPO at $\nu_{\rm QPO} \approx 4.26$ Hz with $Q \approx 14.3$ and a detectable broadband (plateau) noise component, placing it closer to the Type-C regime despite occurring within the same SIMS phase. This gradual transition between QPO types suggests that the Type-B/Type-C classification in this source may represent opposite ends of a single continuous distribution rather than two physically distinct classes.

\subsubsection{Energy and Temporal Evolution of X-ray Lags}
\label{subsubsec:lag_results}

\begin{figure*}
    \centering
    \includegraphics[width=\linewidth,height=9cm]{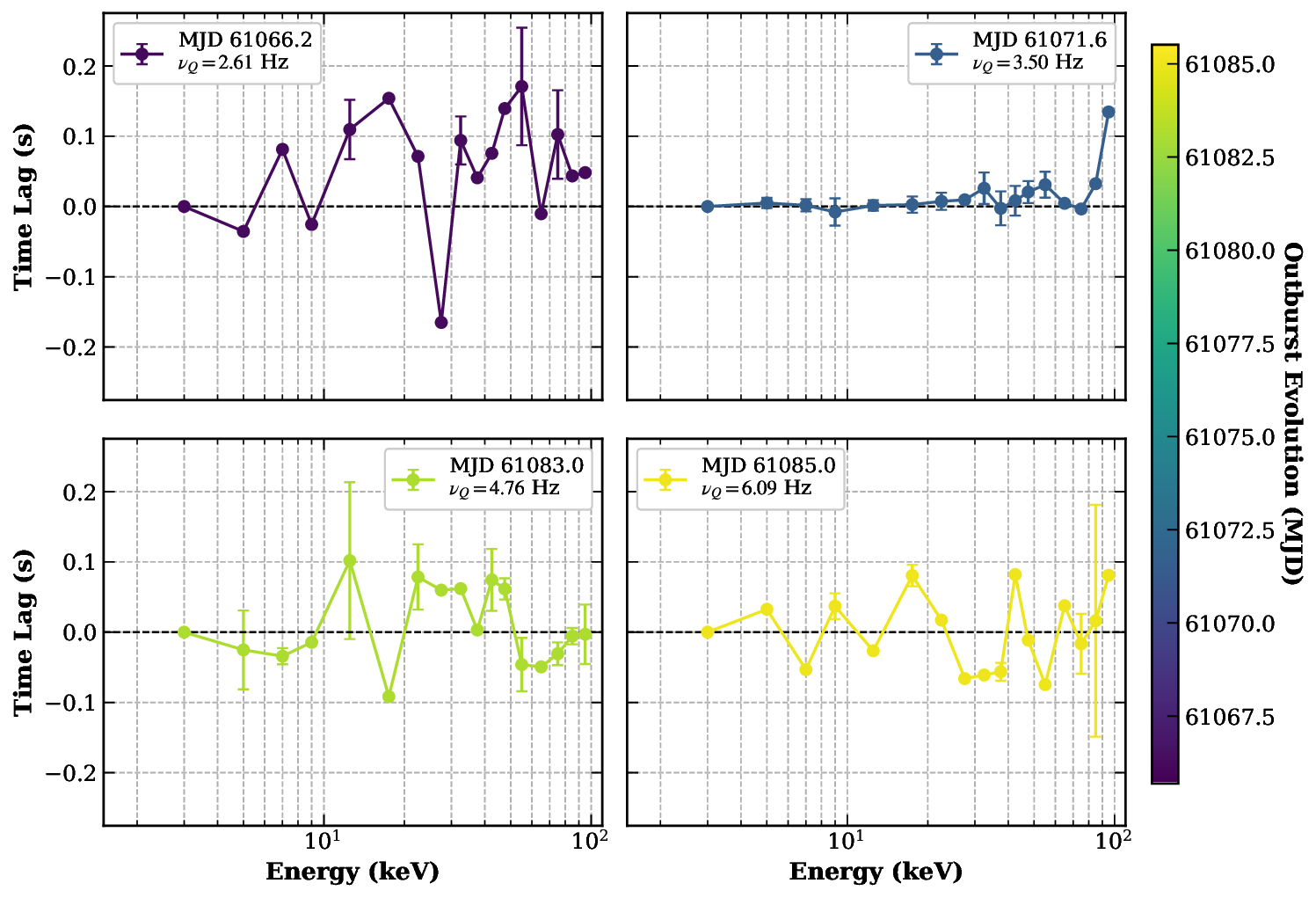}
    \caption{Evolution of the lag-energy spectra for GS 1354-64 across four representative intermediate-state observations ($\nu_{\rm QPO} = 2.61, 3.50, 4.76,$ and $6.09$~Hz). The time lag is computed relative to the 2--4~keV reference band and evaluated at the fundamental QPO frequency. The data are truncated at 100~keV to avoid background dominance. The color map traces the outburst evolution (MJD), demonstrating the systematic collapse of the hard Comptonization delay as the source softens.}
    \label{fig:lag_energy} 
\end{figure*}

\begin{figure*}[!htbp]
    \centering
    \includegraphics[width=\linewidth,height=10cm]{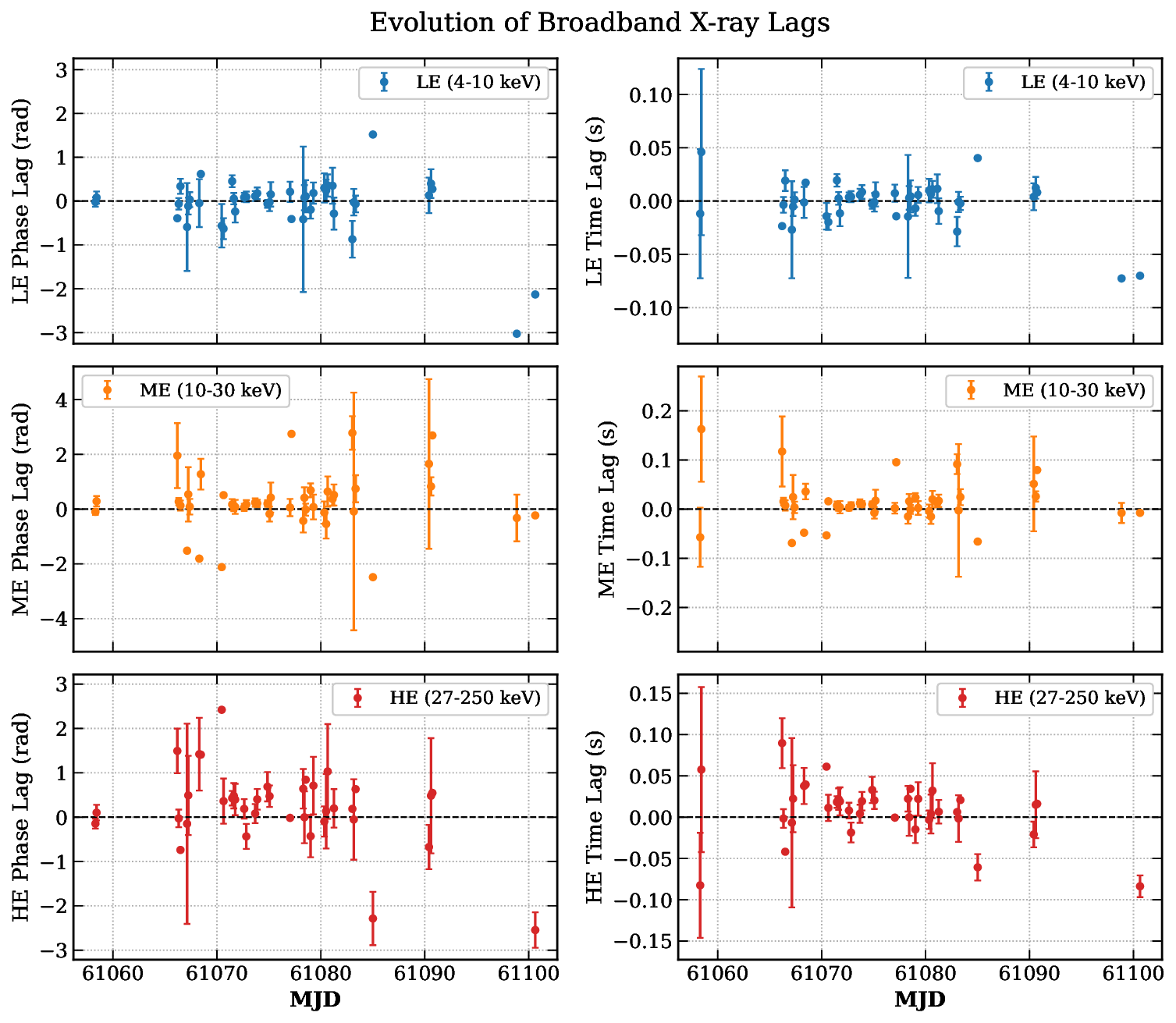}
    \caption{Macroscopic evolution of the broadband X-ray lags over the course of the outburst. \textit{Top:} Phase lag ($\Delta\phi$) in radians. \textit{Bottom:} Absolute time lag ($\tau$) in seconds. The LE (4--10~keV), ME (10--30~keV), and HE (27--250~keV) bands are explicitly independent of the 2--4~keV reference band. A sharp geometric transition is visible near MJD 61070.}
    \label{fig:lag_evolution} 
\end{figure*}

The time delays between the soft, disk-dominated photons and the hard, Comptonized photons provide a geometrical tracer of the accretion flow. In Figure~\ref{fig:lag_energy}, we present the lag-energy evolution, evaluated up to 100~keV to maintain adequate signal-to-noise, for four representative observations spanning the hard-to-soft transition, with fundamental QPO frequencies of 2.61, 3.50, 4.76, and 6.09~Hz.

During the harder, lower-frequency epochs ($\nu_{\rm QPO} = 2.61$ and $3.50$~Hz), we observe positive ``hard'' lags that increase with energy. This indicates that the high-energy photons are delayed relative to the 2--4~keV reference band, a signature of inverse Compton scattering within a physically extended, optically thin corona \citep{Miyamoto1988}. As the outburst progresses and the QPO frequency increases to 4.76 and ultimately 6.09~Hz, the entire lag-energy profile progressively shifts downward and flattens. The high-energy delays diminish significantly, transitioning toward zero and eventually becoming slightly negative (soft lags) at the highest frequencies. This systematic suppression of the hard lag indicates a shift from scattering-dominated to reflection-dominated delays, driven by a contracting coronal geometry.

The temporal evolution of these delays is shown in Figure~\ref{fig:lag_evolution}, which maps the phase and time lags of the LE (4--10~keV), ME (10--30~keV), and HE (27--250~keV) bands across the outburst; all broadband lag measurements are listed in Table~\ref{tab:lag_summary}. The HE band exhibits the most dramatic evolution. Prior to MJD 61070, the HE time lag remains highly positive. Coincident with the rapid steepening of the photon index and the sudden acceleration of the QPO frequency (Section~\ref{subsec:spectral_results}), the HE and ME lags drop simultaneously, converging near zero.

\subsection{Spectro-Temporal Correlations}
\label{sec:correlations}

\begin{figure*}[!h]
    \centering
    \includegraphics[width=\textwidth,height=6.5cm]{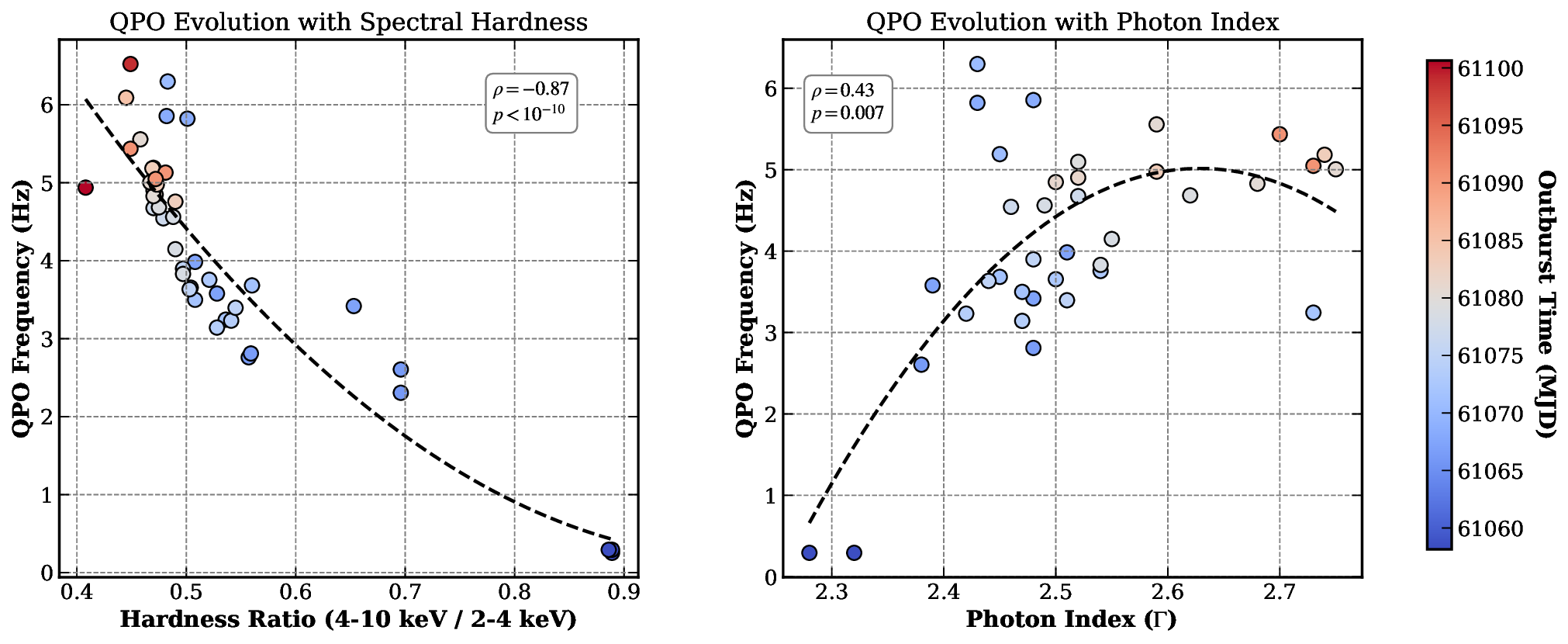}
    \caption{The evolution of the QPO centroid frequency as a function of the spectral hardness ratio (left panel) and the photon index $\Gamma$ (right panel). The data points are color-coded by the observation time (MJD), illustrating the primary state transition and the late-time decay phase. The dashed lines represent simple polynomial trends to guide the eye. Spearman rank correlation coefficients ($\rho$) and their corresponding $p$-values confirm a highly significant spectro-temporal coupling during the outburst.}
    \label{fig:qpo_state_corr}
\end{figure*}

To further quantify the state transition dynamics, we investigated the relationship between the QPO centroid frequency and the concurrent spectral parameters (Figure \ref{fig:qpo_state_corr}). The oscillation frequency exhibits a highly significant anti-correlation with the spectral hardness ($\rho = -0.87$, $p < 10^{-10}$) and a statistically significant moderate positive correlation with the photon index ($\Gamma$; $\rho = 0.43$, $p = 0.007$). These trends, where the QPO frequency rises as the spectrum softens, trace the geometric condensation of the corona during the primary hard-to-soft transition, consistent with phenomenological expectations for black hole transients.

However, we urge caution regarding the QPO behavior during the later, low-luminosity decay phase of the outburst (late-time markers in Figure \ref{fig:qpo_state_corr}). Typically, as a transient decays and the mass accretion rate drops, the inner disk recedes and the QPO frequency is expected to decrease. The persistently high QPO frequencies observed here at low luminosities could suggest a geometric hysteresis, in which the Comptonizing region remains physically compact despite dropping accretion rates. Alternatively, as the source fades, the reduced signal-to-noise ratio complicates the power-spectral modeling, warranting caution when interpreting these late-time detections.

\subsection{Statistical Correlation of Broadband Time Lags}
\label{subsubsec:lag_correlations}

\begin{figure*}[!ht]
    \centering
    \includegraphics[width=\linewidth]{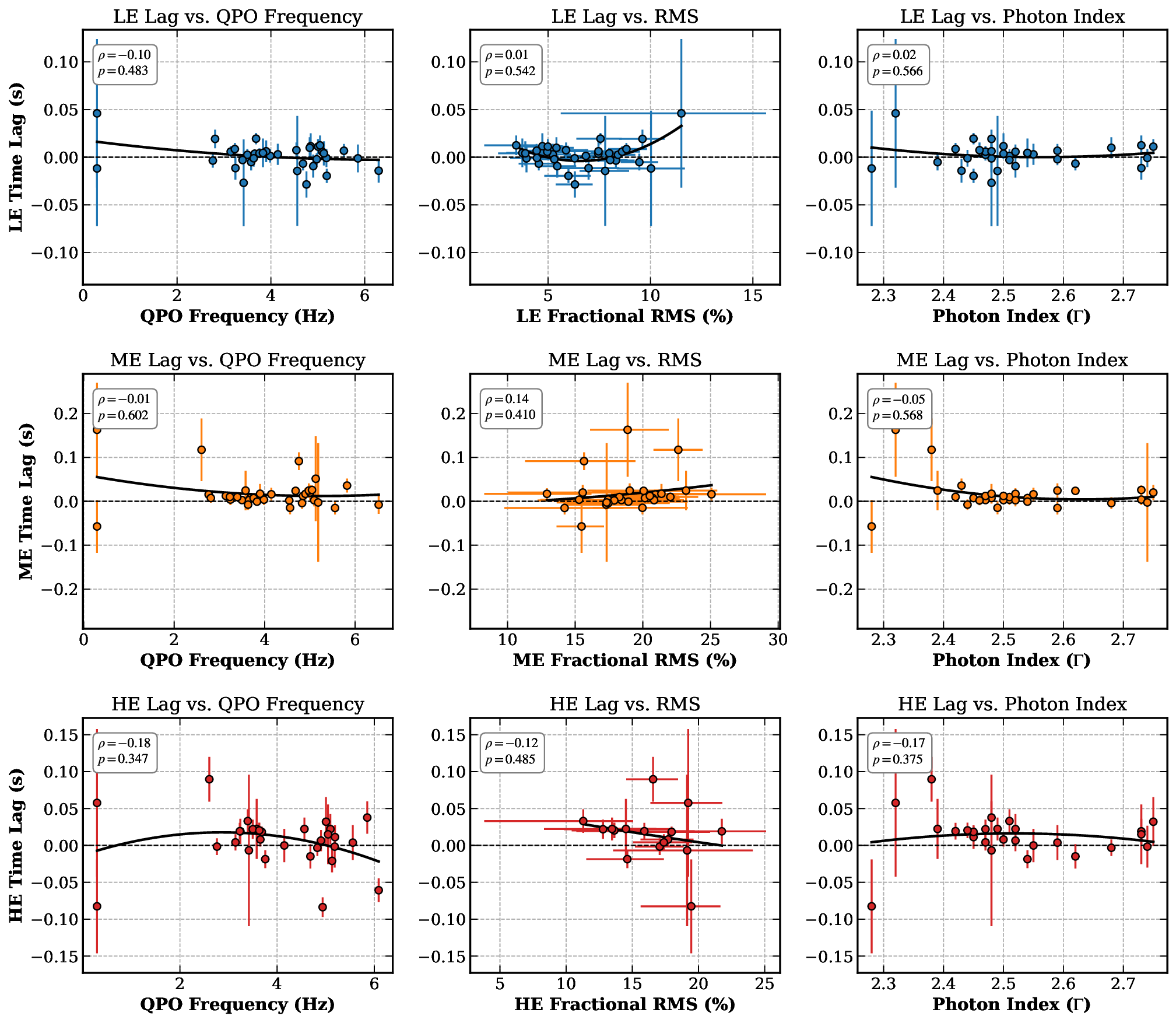}
    \caption{The 3$\times$3 correlation plot, mapping the geometric timing delay (Time Lag) against the global QPO frequency, fractional RMS amplitude, and coronal photon index $\Gamma$ for the LE, ME, and HE energy bands. The statistical significance of each trend is evaluated using a Monte Carlo Spearman rank correlation approach ($10,000$ iterations) to explicitly account for the $1\sigma$ measurement uncertainties. The median correlation coefficients ($\rho$) and $p$-values demonstrate that the large uncertainties during the intermediate states mask any statistically significant global trends ($p > 0.05$).}
    \label{fig:correlations}
\end{figure*}

To quantify the relationship between the time delays and the macroscopic state of the accretion flow, we performed a statistical correlation analysis of the broadband X-ray time lags against the QPO frequency, fractional RMS amplitude, and coronal photon index ($\Gamma$) in the LE, ME, and HE energy bands (Figure~\ref{fig:correlations}). To account for the measurement uncertainties, all reported statistical values were derived with a Monte Carlo resampling approach. Rather than relying on a standard non-parametric test that ignores error bars, we repeatedly perturbed each data point within its $1\sigma$ measurement uncertainties to generate a distribution of 10,000 simulated datasets. The Spearman rank correlation coefficient ($\rho$) and its associated $p$-value were then taken as the medians of these distributions.

The analysis reveals that the soft LE band (4--10~keV) time lags show no significant correlation with any macroscopic state tracer, including the QPO frequency ($\rho = -0.096, p \simeq 0.48$) and the photon index ($\rho = 0.021, p \simeq 0.57$). Similarly, the LE fractional RMS shows no significant correlation ($\rho = 0.010, p \simeq 0.54$). The Medium-Energy (ME; 10--30 keV) band shows a similar lack of correlation across all tested parameters ($p > 0.40$).

With the measurement uncertainties included, the apparent trends in the corona-dominated High-Energy (HE; 27--250~keV) band are also found to lack statistical significance. The HE time lags do not exhibit a robust correlation with the QPO frequency ($\rho = -0.176, p \simeq 0.35$), the photon index ($\rho = -0.174, p \simeq 0.38$), or the fractional RMS amplitude ($\rho = -0.121, p \simeq 0.49$). This indicates that the large measurement uncertainties of the high-energy time lags during the intermediate states obscure any underlying macroscopic correlation.

\section{Discussion}
\label{sec:discussion}

The broadband spectral and temporal monitoring of GS 1354-64 during its 2025--26 outburst provides an opportunity to map the dynamic geometry of the inner accretion flow. Below we discuss the spectral evolution and the disappearance of reflection (Section~\ref{subsec:disc_spectral}), the classification and origin of the QPOs (Section~\ref{subsec:qpo_discussion}), and the time lags (Section~\ref{subsec:geometric_contraction}), which together point to a dynamically contracting corona.

\subsection{Spectral State Transitions and Coronal Cooling}
\label{subsec:disc_spectral}

The outburst of GS 1354-64 exhibited a rapid spectral evolution from a Compton-dominated hard state to a disk-dominated SIMS. During the early phases (prior to MJD 61066), the broadband spectrum was characterized by a relatively flat coronal continuum ($\Gamma \simeq 2.4$), indicating that the inner accretion flow was still dominated by a geometrically thick, optically thin cloud of hot electrons. However, the most critical phase of the outburst occurred near MJD 61070, marked by a rapid steepening of the photon index, peaking at $\Gamma \approx 3.05$.

This rapid spectral softening could be triggered by a sudden surge in the accretion rate of the standard Keplerian disk \citep[in the TCAF framework;][]{Chakrabarti1995, Chakrabarti2015}. As the Keplerian disk migrates inward, it increases the flux of soft seed photons irradiating the hot, sub-Keplerian CENBOL, which acts as the Comptonizing corona. This enhanced inverse-Compton cooling rapidly drains the thermal energy of the CENBOL, leading to the observed steepening of the power-law tail.

A notable aspect of our spectral analysis is the lack of continuous, prominent reflection signatures, such as a strong, relativistically broadened Fe K$\alpha$ emission line or an isolated high-energy Compton hump, in the broadband \textit{Insight}-HXMT spectra across the transition. While our first high-resolution \textit{XRISM} observation captured broad Fe K emission during the early hard state, these features were undetected in the subsequent \textit{XRISM} epoch and remained absent throughout the intermediate states observed by \textit{Insight}-HXMT.

Physically, we interpret this absence of reflection features as a consequence of the rapid geometric evolution and efficient cooling of the corona. As the Keplerian disk surges inward and floods the inner flow with soft seed photons, the suppressed hard X-ray continuum is no longer luminous enough to strongly illuminate the accretion disk, and the fluorescence and back-scattering processes that produce the Fe K$\alpha$ line and Compton hump are severely quenched. Furthermore, as the corona physically contracts, the solid angle it subtends over the accretion disk decreases, starving the disk of illuminating photons and explaining the disappearance of the relativistically broadened reflection features.

This interpretation is supported by our transition to a reflection-free \texttt{diskbb + powerlaw} continuum for the later soft epochs, and is confirmed by our second high-resolution \textit{XRISM} observation (Figure \ref{fig:model_compare}), which revealed a featureless continuum devoid of Fe K$\alpha$ emission. Furthermore, the subsequent persistence of a relatively hot inner disk ($kT_{\rm in} \sim 1.0$~keV) alongside a hardened non-thermal tail indicates that the source entered the SIMS, where both the truncated disk and a compact, cooled corona coexist before the system gradually transitions toward the decline phase.

Finally, it is important to address the large variation in the fitted equivalent hydrogen column density ($N_{\rm H} \sim 0.1$--$2.3 \times 10^{22}$~cm$^{-2}$) across the outburst. While flaring and intermediate accretion states can be accompanied by outflows or disk winds, such a large variation cannot be explained solely by an outflow. In this work, the apparent variation may result from (1) the use of simplified phenomenological spectral models, which can introduce degeneracies between $N_{\rm H}$ and the continuum components at soft energies, and (2) the use of different spectral models (e.g., \texttt{relxill} and \texttt{diskbb+powerlaw}) and fitting parameters for different stages of the outburst as the dominant emission components evolve. Until these methodological effects are carefully assessed and shown to be insufficient to explain the variation, the observed changes should not be interpreted as direct evidence for genuine physical evolution in the source, such as variations in the local absorbing material or the development of an outflow.

\subsection{Classification and Physical Origin of the QPOs}
\label{subsec:qpo_discussion}

The classification of LFQPOs into Types A, B, and C is based on a combination of their centroid frequencies, coherence ($Q$-values), fractional RMS amplitudes, and the underlying broadband noise properties \citep{Casella2005}. Throughout the observed segment of the 2025--2026 outburst of GS 1354-64, the detected oscillations exhibit the classical hallmarks of Type-C QPOs, except for the possible Type-B QPOs over three days between MJD $\sim$ 61075--61078.

The physical origin of Type-C QPOs is widely interpreted within the framework of geometric Lense-Thirring (LT) precession \citep{Stella1998, Ingram2009}. In this paradigm, an inner, hot, geometrically thick accretion flow (the corona) is misaligned with the spin axis of the black hole, and frame dragging forces this entire inner volume to precess as a rigid body. As the mass accretion rate surges during the outburst, the optically thick, geometrically thin accretion disk \citep{Shakura1973} migrates inward and truncates the precessing hot inner flow at progressively smaller radii \citep{Liao2025}. Because the LT precession frequency strongly depends on the outer radius of the precessing flow, the physical shrinking of this geometry, inferred independently from our observed steepening of $\Gamma$ and the simultaneous collapse of the time lags, naturally drives the QPO frequency higher across MJD 61058 to 61098. The LT precession model is also favored by recent phase-resolved spectral studies of Type-C QPOs \citep[e.g.,][]{Ingram2016}.

While LT precession is a leading interpretation for the origin of Type-C QPOs, other mechanisms have been proposed, including accretion-ejection instabilities \citep[e.g.,][]{Tagger1999} and boundary layer oscillations. In particular, the dynamical evolution of the QPOs can alternatively be understood in terms of an oscillating corona, as in the TCAF model \citep{Chakrabarti1995}.

In the TCAF framework, the Comptonizing corona is modeled as a hot, post-shock centrifugal boundary layer (the CENBOL). Within this scenario, LFQPOs are generated by the radial oscillation of the shock front when the local cooling timescale roughly matches the infall timescale. The oscillation of this hot volume modulates the inverse-Compton scattered hard X-ray flux, providing a direct physical mechanism for the amplification of the fractional RMS that we observe in the ME and HE bands. Furthermore, as the outburst progresses, the inward migration of the Keplerian disk increases the supply of soft seed photons, which efficiently cool the CENBOL. This cooling reduces the internal thermal pressure, causing the shock boundary to move closer to the innermost stable circular orbit. A smaller, more compact CENBOL naturally oscillates at a higher frequency \citep{Molteni1996, Chakrabarti2000, Chatterjee2020, Chatterjee2021, Chatterjee2024, Chatterjee2025a}, explaining the observed rise in the QPO frequency from $0.30$~Hz to $6.52$~Hz as the source evolved toward the soft state. Finally, the disappearance of the HE-band QPO at high frequencies may indicate that the Comptonizing region becomes substantially weakened or more compact as the source approaches the disk-dominated soft state.

\subsection{Geometric Evolution and Competing Delay Mechanisms}
\label{subsec:geometric_contraction}

The cross-spectral timing analysis of GS 1354-64 reveals a complex relationship among the spectral softening, the amplitude of the X-ray variability, and the geometric scale of the inner accretion flow. As shown in Section~\ref{subsubsec:lag_correlations}, the broadband X-ray time lags do not follow a simple, strictly monotonic decay, but rather highlight the decoupling of the thermal disk from the corona and the competition between distinct physical delay mechanisms.

As established by the lack of statistical correlation in our Monte Carlo analysis, the soft LE band (4--10~keV) time lags are not driven by the global contraction of the corona. Instead, their relative time delay is likely governed by an interplay of localized inner-disk processes, such as viscous propagation and thermal reprocessing, which evolve as the accretion disk progressively overtakes the power-law emission in these softer bands.

In the corona-dominated HE (27--250~keV) band, the inclusion of the measurement uncertainties reveals that the time lags are also statistically decoupled from the QPO frequency, fractional RMS, and photon index. As the accretion rate surges and the cold disk migrates inward, it floods the corona with seed photons, efficiently cooling it, and this rapid cooling is expected to be accompanied by a collapse of the coronal scale height. However, rather than indicating a monolithic corona that shrinks uniformly and produces a clean decay in the time lags, the scattered temporal behavior suggests that the observed lags are the net result of a tug-of-war between competing mechanisms.

The large uncertainties and lack of monotonic correlation in the high-energy lags reflect this dynamic instability. Long, positive hard lags arise from forward Comptonization delays within an extended coronal geometry, while large, negative soft lags are driven by thermal reverberation as hard X-rays reflect off the dense, migrating accretion disk. As the corona contracts and the inner disk approaches the black hole, these reverberation delays become more prominent, competing with the Comptonization delays and driving the net measured lags toward zero or negative values. Ultimately, while the spectral softening and QPO acceleration provide strong evidence for a dynamically evolving, relativistic accretion geometry (Section~\ref{subsec:qpo_discussion}), the broadband time lags highlight the turbulent, inhomogeneous nature of this state transition.

\section{Summary and Conclusions}
\label{sec:summary}

In this work, we have used the combined spectroscopic and high-cadence timing capabilities of \textit{XRISM} and \textit{Insight}-HXMT to trace the geometric and physical evolution of the Galactic black hole transient GS 1354-64 during its 2025--2026 outburst. By combining the broadband continuum behavior, high-resolution reflection features, low-frequency QPOs, and energy-dependent X-ray lags, we arrive at the following conclusions:

\begin{enumerate}
    \item \textbf{Complete State Transition:} Unlike its 1997 and 2015 outbursts, which remained in the hard state, GS 1354-64 evolved through a complete hard-to-soft state transition. The broadband continuum shifted from a heavily Comptonized spectrum ($\Gamma \sim 2.28$) to a thermally dominated Soft-Intermediate State ($\Gamma \sim 2.77$, $kT_{\rm in} \sim 1.16$~keV).

    \item \textbf{Quenching of Relativistic Reflection:} High-resolution \textit{XRISM}-Resolve spectroscopy demonstrates a clear spectral pivoting. The broad Fe K$\alpha$ emission and reflection continuum observed during the hard-intermediate state were quenched in the softer epochs, indicating that the efficient cooling and physical contraction of the corona starved the accretion disk of the hard X-ray irradiation required to sustain relativistic reflection.

    \item \textbf{Dynamical QPO Evolution:} We detected Type-C QPOs evolving from $0.30$~Hz to $6.52$~Hz. The oscillation frequency is anti-correlated with the spectral hardness and positively correlated with the photon index. The high fractional RMS amplitude of these QPOs in the ME and HE bands shows that the variability is governed by the hot coronal flow rather than the thermal disk.

    \item \textbf{Collapse of the X-ray Time Lags:} Cross-spectral analysis revealed large positive time lags during the low-frequency epochs, indicative of inverse-Compton scattering in an extended corona. As the QPO frequency increased beyond $\sim 3$--$4$~Hz, the high-energy hard lags collapsed toward zero. The LE band (4--10~keV) lags remained flat and statistically decoupled from the QPO evolution, suggesting that the soft-band delays are driven by localized intra-disk propagation rather than coronal scattering.

    \item \textbf{Geometric Contraction and Competing Lags:} While both an inwardly oscillating shock front (TCAF) and a precessing misaligned hot flow (LT) can produce higher oscillation frequencies as the geometry shrinks, the LT precession framework is favored by the available literature \citep[e.g.,][]{Ingram2016}. However, our Monte Carlo correlation analysis shows that the broadband time lags lack a statistically significant monotonic trend, highlighting the inhomogeneous nature of the inner flow during the state transition, in which positive Comptonization delays and negative reverberation delays compete as the corona contracts.
\end{enumerate}

Ultimately, the 2025--26 outburst of GS 1354-64 provides a benchmark for accretion physics, linking the thermodynamic cooling of the corona to its macroscopic geometric contraction.

\section*{Data Availability}
\label{sec:data_avail}

The \textit{Insight}-HXMT data used in this work (proposal IDs P0814047 and P0814048) are publicly available from the \textit{Insight}-HXMT archive, and the \textit{XRISM} data (ObsIDs 902002010 and 902002020) from the HEASARC and DARTS/ISAS archives. 

This work has made use of public data from several satellite/instrument archives and has made use of software from the HEASARC, which is developed and monitored by the Astrophysics Science Division at NASA/GSFC and the High Energy Astrophysics Division of the Smithsonian Astrophysical Observatory. This work made use of the data from the {\it Insight}-HXMT mission, a project funded by the China National Space Administration (CNSA) and the Chinese Academy of Sciences (CAS).

The data reduction pipelines, calibration databases (CALDB), and software used in this work, namely HEASoft \texttt{FTOOLS} and the \texttt{HXMTDAS} package, are publicly available from their respective official repositories.

\section*{Acknowledgements}
\label{sec:ack}

K. C., B. K., \& X.-W. L. acknowledge support from the ``Science \& Technology Champion Project'' (202005AB160002) and from two ``Team Projects'' -- the ``Top Team'' (202305AT350002) and the ``Innovation Team'' (202105AE160021), all funded by the ``Yunnan Revitalization Talent Support Program''. They also acknowledge the support from the ``Key Laboratory of Survey Science of Yunnan Province" with project No. 202449CE340002. KC also acknowledges support from the 2026 National Natural Science Foundation of China (NSFC) RFIS-I Grant (W2633019), the 79th General Grant the China Postdoctoral Science Foundation (Certificate Number: 2026M793513), and CaiYun-2024 Postdoctoral Grant for Innovation Project of Yunnan Province (C615300504128). K.C. also acknowledges additional support from the ``High-Level Talent Introduction Program Funding'' of Prof. Jianhui Lian of SWIFAR, Yunnan University (W820301911). B.P. acknowledges collective support from Narodowe Centrum Nauki (NCN) grants 2021/41/B/ST9/04110 and 2018/31/G/ST9/03224 for this research. A.T. acknowledges support from the Tianshan Talent Training Program (grant No. 2023TSYCCX0099) and the Xinjiang Tianchi Talent Program. B. K. also acknowledges the support from the `Special Project for High-End Foreign Experts, Xingdian Funding from Yunnan Province, and the National Key Research and Development Program of China (2024YFA1611603). H.-K. C. acknowledges support from the NSTC project (grant No. 114-2112-M-007-042) of NTHU.

We thank the anonymous referee for making fruitful comments which have made the manuscript scientifically better.

\section*{Facilities and Softwares}
\label{sec:facilities}

\textbf{Facilities:} \textit{XRISM} (JAXA/NASA/ESA), \textit{Insight}-HXMT (CAS).

\textbf{Software:} This research made use of the \texttt{HEASoft} suite (v6.35) of data reduction and analysis tools, including \texttt{FTOOLS} and \texttt{XSPEC} (v12.15.1) \citep{Arnaud1996}, provided by the High Energy Astrophysics Science Archive Research Center (HEASARC). Graphical visualizations were executed in a \texttt{Python} (v3.10) environment utilizing the \texttt{NumPy}, \texttt{SciPy}, and \texttt{Matplotlib} libraries.

\begin{center}
\makeatletter\def\table@hook{\scriptsize}\makeatother 
\scriptsize
\begin{longtable}{llccccc}
\caption{Comprehensive Summary of Observations} \label{tab:obs_log} \\
\hline\hline
Instrument & ObsID / Exposure ID & Start UT & End UT & Start MJD & End MJD & Avg MJD \\
\hline
\endfirsthead
\multicolumn{7}{c}%
{\tablename\ \thetable\ -- \textit{Continued from previous page}} \\
\hline\hline
Instrument & ObsID / Exposure ID & Start UT & End UT & Start MJD & End MJD & Avg MJD \\
\hline
\endhead
\hline \multicolumn{7}{r}{\textit{Continued on next page}} \\
\endfoot
\hline\hline
\endlastfoot

\multicolumn{7}{l}{\textbf{XRISM}} \\
\hline
XRISM & 902002010 & 2026-01-20 03:38:04 & 2026-01-23 03:27:40 & 61060.15141 & 61063.14421 & 61061.64781 \\
XRISM & 902002020 & 2026-02-19 17:25:04 & 2026-02-20 10:09:32 & 61090.72574 & 61091.42328 & 61091.07451 \\
\hline
\multicolumn{7}{l}{\textbf{Insight-HXMT}} \\
\multicolumn{7}{l}{\textit{\ \ Proposal ID: P0814047}} \\
HXMT & P081404700101-20260118-01-01 & 2026-01-18 02:25:42 & 2026-01-18 05:58:30 & 61058.10118 & 61058.24896 & 61058.17507 \\
HXMT & P081404700102-20260118-01-01 & 2026-01-18 05:58:30 & 2026-01-18 09:24:01 & 61058.24896 & 61058.39168 & 61058.32032 \\
HXMT & P081404700103-20260118-01-01 & 2026-01-18 09:24:01 & 2026-01-18 11:55:17 & 61058.39168 & 61058.49672 & 61058.44420 \\
HXMT & P081404700201-20260127-01-01 & 2026-01-27 01:40:28 & 2026-01-27 05:08:45 & 61067.06977 & 61067.21441 & 61067.14209 \\
HXMT & P081404700202-20260127-01-01 & 2026-01-27 05:08:45 & 2026-01-27 08:29:19 & 61067.21441 & 61067.35369 & 61067.28405 \\
HXMT & P081404700203-20260127-01-01 & 2026-01-27 08:29:19 & 2026-01-27 11:15:08 & 61067.35369 & 61067.46884 & 61067.41127 \\
HXMT & P081404700401-20260311-01-01 & 2026-03-11 01:06:35 & 2026-03-11 04:37:06 & 61110.04624 & 61110.19243 & 61110.11933 \\
HXMT & P081404700402-20260311-01-01 & 2026-03-11 04:37:06 & 2026-03-11 07:50:31 & 61110.19243 & 61110.32675 & 61110.25959 \\
HXMT & P081404700403-20260311-01-01 & 2026-03-11 07:50:31 & 2026-03-11 10:49:10 & 61110.32675 & 61110.45081 & 61110.38878 \\
\multicolumn{7}{l}{\textit{\ \ Proposal ID: P0814048}} \\
HXMT & P081404800101-20260126-01-01 & 2026-01-26 00:20:03 & 2026-01-26 03:39:31 & 61066.01392 & 61066.15244 & 61066.08318 \\
HXMT & P081404800102-20260126-01-01 & 2026-01-26 03:39:31 & 2026-01-26 07:07:11 & 61066.15244 & 61066.29666 & 61066.22455 \\
HXMT & P081404800103-20260126-01-01 & 2026-01-26 07:07:11 & 2026-01-26 10:25:56 & 61066.29666 & 61066.43468 & 61066.36567 \\
HXMT & P081404800104-20260126-01-01 & 2026-01-26 10:25:56 & 2026-01-26 14:39:39 & 61066.43468 & 61066.61087 & 61066.52277 \\
HXMT & P081404800301-20260128-01-01 & 2026-01-28 06:11:25 & 2026-01-28 09:51:15 & 61068.25793 & 61068.41059 & 61068.33426 \\
HXMT & P081404800302-20260128-01-01 & 2026-01-28 09:51:15 & 2026-01-28 13:29:48 & 61068.41059 & 61068.56236 & 61068.48648 \\
HXMT & P081404800303-20260128-01-01 & 2026-01-28 13:29:48 & 2026-01-28 15:46:15 & 61068.56236 & 61068.65712 & 61068.60974 \\
HXMT & P081404800601-20260129-01-01 & 2026-01-29 05:57:29 & 2026-01-29 09:33:54 & 61069.24825 & 61069.39854 & 61069.32340 \\
HXMT & P081404800602-20260129-01-01 & 2026-01-29 09:33:54 & 2026-01-29 13:12:42 & 61069.39854 & 61069.55049 & 61069.47451 \\
HXMT & P081404800603-20260129-01-01 & 2026-01-29 13:12:42 & 2026-01-29 15:32:22 & 61069.55049 & 61069.64748 & 61069.59898 \\
HXMT & P081404800701-20260130-01-01 & 2026-01-30 10:28:45 & 2026-01-30 14:30:28 & 61070.43663 & 61070.60449 & 61070.52056 \\
HXMT & P081404800702-20260130-01-01 & 2026-01-30 14:30:28 & 2026-01-30 17:40:12 & 61070.60449 & 61070.73625 & 61070.67037 \\
HXMT & P081404800703-20260130-01-01 & 2026-01-30 17:40:12 & 2026-01-30 20:03:30 & 61070.73625 & 61070.83576 & 61070.78601 \\
HXMT & P081404800801-20260131-01-01 & 2026-01-31 10:14:24 & 2026-01-31 14:13:22 & 61071.42667 & 61071.59262 & 61071.50964 \\
HXMT & P081404800802-20260131-01-01 & 2026-01-31 14:13:22 & 2026-01-31 17:23:06 & 61071.59262 & 61071.72437 & 61071.65850 \\
HXMT & P081404800803-20260131-01-01 & 2026-01-31 17:23:06 & 2026-01-31 19:48:54 & 61071.72437 & 61071.82562 & 61071.77500 \\
HXMT & P081404800901-20260201-01-01 & 2026-02-01 14:44:24 & 2026-02-01 18:40:51 & 61072.61417 & 61072.77837 & 61072.69627 \\
HXMT & P081404800902-20260201-01-01 & 2026-02-01 18:40:51 & 2026-02-01 21:50:34 & 61072.77837 & 61072.91012 & 61072.84424 \\
HXMT & P081404801001-20260202-01-01 & 2026-02-02 16:03:37 & 2026-02-02 19:58:32 & 61073.66918 & 61073.83231 & 61073.75075 \\
HXMT & P081404801002-20260202-01-01 & 2026-02-02 19:58:32 & 2026-02-02 23:08:15 & 61073.83231 & 61073.96406 & 61073.89819 \\
HXMT & P081404801101-20260203-01-01 & 2026-02-03 20:32:16 & 2026-02-04 00:25:58 & 61074.85574 & 61075.01803 & 61074.93689 \\
HXMT & P081404801102-20260204-02-01 & 2026-02-04 00:25:58 & 2026-02-04 03:35:41 & 61075.01803 & 61075.14978 & 61075.08391 \\
HXMT & P081404801103-20260204-02-01 & 2026-02-04 03:35:41 & 2026-02-04 06:16:10 & 61075.14978 & 61075.26123 & 61075.20550 \\
HXMT & P081404801301-20260206-01-01 & 2026-02-06 00:43:50 & 2026-02-06 03:56:07 & 61077.03044 & 61077.16397 & 61077.09720 \\
HXMT & P081404801302-20260206-01-01 & 2026-02-06 03:56:07 & 2026-02-06 07:15:01 & 61077.16397 & 61077.30209 & 61077.23303 \\
HXMT & P081404801303-20260206-01-01 & 2026-02-06 07:15:01 & 2026-02-06 10:29:08 & 61077.30209 & 61077.43690 & 61077.36950 \\
HXMT & P081404801401-20260207-01-01 & 2026-02-07 05:11:43 & 2026-02-07 09:03:41 & 61078.21647 & 61078.37756 & 61078.29701 \\
HXMT & P081404801402-20260207-01-01 & 2026-02-07 09:03:41 & 2026-02-07 12:13:24 & 61078.37756 & 61078.50931 & 61078.44343 \\
HXMT & P081404801403-20260207-01-01 & 2026-02-07 12:13:24 & 2026-02-07 14:45:25 & 61078.50931 & 61078.61487 & 61078.56209 \\
HXMT & P081404801501-20260210-01-01 & 2026-02-10 02:46:11 & 2026-02-10 06:05:22 & 61081.11541 & 61081.25373 & 61081.18457 \\
HXMT & P081404801502-20260210-01-01 & 2026-02-10 06:05:22 & 2026-02-10 09:47:17 & 61081.25373 & 61081.40784 & 61081.33078 \\
HXMT & P081404801503-20260210-01-01 & 2026-02-10 09:47:17 & 2026-02-10 12:19:51 & 61081.40784 & 61081.51378 & 61081.46081 \\
HXMT & P081404801601-20260318-01-01 & 2026-03-18 14:56:57 & 2026-03-18 19:08:21 & 61117.62288 & 61117.79747 & 61117.71017 \\
HXMT & P081404801602-20260318-01-01 & 2026-03-18 19:08:21 & 2026-03-18 22:17:59 & 61117.79747 & 61117.92916 & 61117.86331 \\
HXMT & P081404801603-20260318-01-01 & 2026-03-18 22:17:59 & 2026-03-19 00:32:03 & 61117.92916 & 61118.02226 & 61117.97571 \\
HXMT & P081404801901-20260208-01-01 & 2026-02-08 00:10:14 & 2026-02-08 03:21:08 & 61079.00711 & 61079.13968 & 61079.07339 \\
HXMT & P081404801903-20260208-01-01 & 2026-02-08 06:40:12 & 2026-02-08 09:43:54 & 61079.27792 & 61079.40549 & 61079.34170 \\
HXMT & P081404802001-20260209-01-01 & 2026-02-09 07:47:43 & 2026-02-09 11:39:11 & 61080.32480 & 61080.48554 & 61080.40517 \\
HXMT & P081404802002-20260209-01-01 & 2026-02-09 11:39:11 & 2026-02-09 14:48:54 & 61080.48554 & 61080.61729 & 61080.55142 \\
HXMT & P081404802003-20260209-01-01 & 2026-02-09 14:48:54 & 2026-02-09 17:21:24 & 61080.61729 & 61080.72319 & 61080.67024 \\
HXMT & P081404802101-20260212-01-01 & 2026-02-12 00:37:30 & 2026-02-12 03:50:26 & 61083.02604 & 61083.16002 & 61083.09303 \\
HXMT & P081404802102-20260212-01-01 & 2026-02-12 03:50:26 & 2026-02-12 07:38:48 & 61083.16002 & 61083.31861 & 61083.23932 \\
HXMT & P081404802103-20260212-01-01 & 2026-02-12 07:38:48 & 2026-02-12 11:54:27 & 61083.31861 & 61083.49615 & 61083.40738 \\
HXMT & P081404802201-20260214-01-01 & 2026-02-14 00:03:41 & 2026-02-14 05:01:09 & 61085.00256 & 61085.20913 & 61085.10584 \\
HXMT & P081404802301-20260215-01-01 & 2026-02-15 18:45:24 & 2026-02-15 23:21:31 & 61086.78153 & 61086.97328 & 61086.87740 \\
HXMT & P081404802302-20260215-01-01 & 2026-02-15 23:21:31 & 2026-02-16 04:33:24 & 61086.97328 & 61087.18986 & 61087.08157 \\
HXMT & P081404802501-20260217-01-01 & 2026-02-17 15:02:10 & 2026-02-17 19:24:56 & 61088.62650 & 61088.80898 & 61088.71774 \\
HXMT & P081404802502-20260217-01-01 & 2026-02-17 19:24:56 & 2026-02-18 00:50:08 & 61088.80898 & 61089.03481 & 61089.92190 \\
HXMT & P081404802601-20260219-01-01 & 2026-02-19 09:44:13 & 2026-02-19 13:49:29 & 61090.40571 & 61090.57603 & 61090.49087 \\
HXMT & P081404802602-20260219-01-01 & 2026-02-19 13:49:29 & 2026-02-19 17:03:50 & 61090.57603 & 61090.71100 & 61090.64351 \\
HXMT & P081404802603-20260219-01-01 & 2026-02-19 17:03:50 & 2026-02-19 19:28:45 & 61090.71100 & 61090.81163 & 61090.76131 \\
HXMT & P081404802701-20260221-01-01 & 2026-02-21 04:26:34 & 2026-02-21 10:11:45 & 61092.18512 & 61092.42483 & 61092.30497 \\
HXMT & P081404802702-20260221-01-01 & 2026-02-21 10:11:45 & 2026-02-21 14:01:44 & 61092.42483 & 61092.58454 & 61092.50468 \\
HXMT & P081404802801-20260223-01-01 & 2026-02-23 22:53:05 & 2026-02-24 02:16:52 & 61094.95353 & 61095.09505 & 61095.02429 \\
HXMT & P081404802802-20260224-02-01 & 2026-02-24 02:16:52 & 2026-02-24 05:25:31 & 61095.09505 & 61095.22605 & 61095.16055 \\
HXMT & P081404802803-20260224-02-01 & 2026-02-24 05:25:31 & 2026-02-24 08:38:26 & 61095.22605 & 61095.36002 & 61095.29304 \\
HXMT & P081404802901-20260225-01-01 & 2026-02-25 06:31:41 & 2026-02-25 09:47:06 & 61096.27200 & 61096.40771 & 61096.33986 \\
HXMT & P081404802902-20260225-01-01 & 2026-02-25 09:47:06 & 2026-02-25 12:56:20 & 61096.40771 & 61096.53912 & 61096.47341 \\
HXMT & P081404802903-20260225-01-01 & 2026-02-25 12:56:20 & 2026-02-25 16:18:20 & 61096.53912 & 61096.67940 & 61096.60926 \\
HXMT & P081404803001-20260226-01-01 & 2026-02-26 20:30:04 & 2026-02-26 23:38:36 & 61097.85421 & 61097.98514 & 61097.91968 \\
HXMT & P081404803002-20260226-01-01 & 2026-02-26 23:38:36 & 2026-02-27 02:47:58 & 61097.98514 & 61098.11664 & 61098.05089 \\
HXMT & P081404803003-20260227-02-01 & 2026-02-27 02:47:58 & 2026-02-27 06:15:08 & 61098.11664 & 61098.26051 & 61098.18858 \\
HXMT & P081404803101-20260227-01-01 & 2026-02-27 20:14:06 & 2026-02-27 23:19:02 & 61098.84312 & 61098.97155 & 61098.90734 \\
HXMT & P081404803102-20260227-01-01 & 2026-02-27 23:19:02 & 2026-02-28 02:28:26 & 61098.97155 & 61099.10308 & 61099.03731 \\
HXMT & P081404803103-20260228-02-01 & 2026-02-28 02:28:26 & 2026-02-28 06:00:27 & 61099.10308 & 61099.25031 & 61099.17670 \\
HXMT & P081404803201-20260228-01-01 & 2026-02-28 19:58:03 & 2026-02-28 22:59:46 & 61099.83198 & 61099.95817 & 61099.89508 \\
HXMT & P081404803202-20260228-01-01 & 2026-02-28 22:59:46 & 2026-03-01 02:09:14 & 61099.95817 & 61100.08975 & 61100.02396 \\
HXMT & P081404803203-20260301-02-01 & 2026-03-01 02:09:14 & 2026-03-01 05:45:08 & 61100.08975 & 61100.23968 & 61100.16471 \\
HXMT & P081404803301-20260301-01-01 & 2026-03-01 14:57:09 & 2026-03-01 19:13:55 & 61100.62302 & 61100.80133 & 61100.71218 \\
HXMT & P081404803302-20260301-01-01 & 2026-03-01 19:13:55 & 2026-03-02 00:44:19 & 61100.80133 & 61101.03078 & 61100.91605 \\
HXMT & P081404803401-20260304-01-01 & 2026-03-04 03:03:55 & 2026-03-04 06:44:03 & 61103.12772 & 61103.28059 & 61103.20416 \\
HXMT & P081404803402-20260304-01-01 & 2026-03-04 06:44:03 & 2026-03-04 09:56:41 & 61103.28059 & 61103.41436 & 61103.34748 \\
HXMT & P081404803601-20260302-01-01 & 2026-03-02 09:56:12 & 2026-03-02 13:55:53 & 61101.41403 & 61101.58047 & 61101.49725 \\
HXMT & P081404803602-20260302-01-01 & 2026-03-02 13:55:53 & 2026-03-02 17:16:56 & 61101.58047 & 61101.72009 & 61101.65028 \\
HXMT & P081404803603-20260302-01-01 & 2026-03-02 17:16:56 & 2026-03-02 19:41:20 & 61101.72009 & 61101.82037 & 61101.77023 \\
HXMT & P081404803701-20260303-01-01 & 2026-03-03 08:05:02 & 2026-03-03 11:52:11 & 61102.33683 & 61102.49457 & 61102.41570 \\
HXMT & P081404803702-20260303-01-01 & 2026-03-03 11:52:11 & 2026-03-03 15:19:43 & 61102.49457 & 61102.63869 & 61102.56663 \\
HXMT & P081404803703-20260303-01-01 & 2026-03-03 15:19:43 & 2026-03-03 17:38:12 & 61102.63869 & 61102.73486 & 61102.68678 \\
HXMT & P081404803801-20260306-01-01 & 2026-03-06 19:54:32 & 2026-03-07 00:18:03 & 61105.82954 & 61106.01253 & 61105.92104 \\
HXMT & P081404803802-20260307-02-01 & 2026-03-07 00:18:03 & 2026-03-07 05:27:36 & 61106.01253 & 61106.22750 & 61106.12002 \\
HXMT & P081404803901-20260307-01-01 & 2026-03-07 14:53:09 & 2026-03-07 19:07:14 & 61106.62024 & 61106.79669 & 61106.70847 \\
HXMT & P081404803902-20260307-01-01 & 2026-03-07 19:07:14 & 2026-03-08 00:41:33 & 61106.79669 & 61107.02885 & 61106.91277 \\
HXMT & P081404804001-20260308-01-01 & 2026-03-08 14:36:20 & 2026-03-08 18:49:29 & 61107.60856 & 61107.78436 & 61107.69646 \\
HXMT & P081404804002-20260308-01-01 & 2026-03-08 18:49:29 & 2026-03-09 00:21:07 & 61107.78436 & 61108.01466 & 61107.89951 \\
HXMT & P081404804101-20260309-01-01 & 2026-03-09 12:44:34 & 2026-03-09 16:51:36 & 61108.53095 & 61108.70250 & 61108.61672 \\
HXMT & P081404804102-20260309-01-01 & 2026-03-09 16:51:36 & 2026-03-09 20:14:15 & 61108.70250 & 61108.84323 & 61108.77286 \\
HXMT & P081404804103-20260309-01-01 & 2026-03-09 20:14:15 & 2026-03-10 01:36:57 & 61108.84323 & 61109.06733 & 61108.95528 \\
HXMT & P081404804201-20260310-01-01 & 2026-03-10 15:37:20 & 2026-03-10 19:56:13 & 61109.65093 & 61109.83071 & 61109.74082 \\
HXMT & P081404804202-20260310-01-01 & 2026-03-10 19:56:13 & 2026-03-11 01:06:37 & 61109.83071 & 61110.04626 & 61109.93848 \\
HXMT & P081404804301-20260311-01-01 & 2026-03-11 20:04:59 & 2026-03-12 00:22:43 & 61110.83679 & 61111.01578 & 61110.92628 \\
HXMT & P081404804302-20260312-02-01 & 2026-03-12 00:22:43 & 2026-03-12 05:34:14 & 61111.01578 & 61111.23211 & 61111.12394 \\
HXMT & P081404804401-20260312-01-01 & 2026-03-12 05:34:12 & 2026-03-12 09:17:15 & 61111.23208 & 61111.38698 & 61111.30953 \\
HXMT & P081404804402-20260312-01-01 & 2026-03-12 09:17:15 & 2026-03-12 12:39:20 & 61111.38698 & 61111.52731 & 61111.45715 \\
HXMT & P081404804403-20260312-01-01 & 2026-03-12 12:39:20 & 2026-03-12 15:16:14 & 61111.52731 & 61111.63627 & 61111.58179 \\
HXMT & P081404804501-20260313-01-01 & 2026-03-13 16:21:18 & 2026-03-13 20:37:21 & 61112.68146 & 61112.85927 & 61112.77036 \\
HXMT & P081404804502-20260313-01-01 & 2026-03-13 20:37:21 & 2026-03-13 23:46:58 & 61112.85927 & 61112.99095 & 61112.92511 \\
HXMT & P081404804503-20260313-01-01 & 2026-03-13 23:46:58 & 2026-03-14 03:40:48 & 61112.99095 & 61113.15333 & 61113.07214 \\
HXMT & P081404804601-20260314-01-01 & 2026-03-14 16:04:20 & 2026-03-14 20:19:30 & 61113.66968 & 61113.84688 & 61113.75828 \\
HXMT & P081404804602-20260314-01-01 & 2026-03-14 20:19:30 & 2026-03-15 01:48:24 & 61113.84688 & 61114.07528 & 61113.96108 \\
HXMT & P081404804701-20260316-01-01 & 2026-03-16 10:45:53 & 2026-03-16 14:46:27 & 61115.44853 & 61115.61559 & 61115.53206 \\
HXMT & P081404804702-20260316-01-01 & 2026-03-16 14:46:27 & 2026-03-16 18:09:04 & 61115.61559 & 61115.75630 & 61115.68594 \\
HXMT & P081404804703-20260316-01-01 & 2026-03-16 18:09:04 & 2026-03-16 20:25:47 & 61115.75630 & 61115.85124 & 61115.80377 \\
HXMT & P081404804801-20260317-01-01 & 2026-03-17 05:44:24 & 2026-03-17 09:30:13 & 61116.23917 & 61116.39598 & 61116.31758 \\
HXMT & P081404804802-20260317-01-01 & 2026-03-17 09:30:13 & 2026-03-17 12:48:47 & 61116.39598 & 61116.53388 & 61116.46493 \\
HXMT & P081404804901-20260320-01-01 & 2026-03-20 15:58:42 & 2026-03-20 20:53:11 & 61119.66576 & 61119.87027 & 61119.76802 \\
HXMT & P081404805001-20260322-01-01 & 2026-03-22 09:06:19 & 2026-03-22 12:59:02 & 61121.37939 & 61121.54100 & 61121.46019 \\
HXMT & P081404805002-20260322-01-01 & 2026-03-22 12:59:02 & 2026-03-22 16:22:38 & 61121.54100 & 61121.68238 & 61121.61169 \\
HXMT & P081404805003-20260322-01-01 & 2026-03-22 16:22:38 & 2026-03-22 18:47:39 & 61121.68238 & 61121.78309 & 61121.73274 \\
HXMT & P081404805101-20260323-01-01 & 2026-03-23 08:50:04 & 2026-03-23 12:41:09 & 61122.36810 & 61122.52858 & 61122.44834 \\
HXMT & P081404805102-20260323-01-01 & 2026-03-23 12:41:09 & 2026-03-23 16:04:55 & 61122.52858 & 61122.67008 & 61122.59933 \\
HXMT & P081404805103-20260323-01-01 & 2026-03-23 16:04:55 & 2026-03-23 18:30:25 & 61122.67008 & 61122.77112 & 61122.72060 \\
HXMT & P081404805501-20260328-01-01 & 2026-03-28 18:33:32 & 2026-03-28 22:30:16 & 61127.77329 & 61127.93769 & 61127.85549 \\
HXMT & P081404805502-20260328-01-01 & 2026-03-28 22:30:16 & 2026-03-29 01:39:54 & 61127.93769 & 61128.06937 & 61128.00353 \\
HXMT & P081404805503-20260329-02-01 & 2026-03-29 01:39:54 & 2026-03-29 04:13:49 & 61128.06937 & 61128.17626 & 61128.12282 \\
HXMT & P081404805601-20260330-01-01 & 2026-03-30 13:15:42 & 2026-03-30 17:10:16 & 61129.55257 & 61129.71546 & 61129.63402 \\
HXMT & P081404805602-20260330-01-01 & 2026-03-30 17:10:16 & 2026-03-30 20:19:53 & 61129.71546 & 61129.84714 & 61129.78130 \\
HXMT & P081404805603-20260330-01-01 & 2026-03-30 20:19:53 & 2026-03-30 22:55:55 & 61129.84714 & 61129.95550 & 61129.90132 \\
HXMT & P081404805701-20260401-01-01 & 2026-04-01 07:57:20 & 2026-04-01 11:50:10 & 61131.33148 & 61131.49317 & 61131.41233 \\
HXMT & P081404805702-20260401-01-01 & 2026-04-01 11:50:10 & 2026-04-01 14:59:47 & 61131.49317 & 61131.62485 & 61131.55901 \\
HXMT & P081404805703-20260401-01-01 & 2026-04-01 14:59:47 & 2026-04-01 17:37:31 & 61131.62485 & 61131.73439 & 61131.67962 \\
\end{longtable}
\normalsize
\end{center}

\startlongtable
\begin{deluxetable*}{l | c | c | c | c | c | c | c | c | c | c | c}
\tabletypesize{\scriptsize}
\setlength{\tabcolsep}{3pt}
\tablewidth{0pt}
\tablecaption{Summary of accepted broadband spectral fits for GS~1354$-$64 from HXMT. \label{tab:spec_model_summary}}

\tablehead{
    \colhead{ObsID / Exposure ID} & \colhead{Model} & \colhead{$C_{\rm ME}$} & \colhead{$C_{\rm HE}$} & \colhead{$N_{\rm H}$} & \colhead{$kT_{\rm in}$} & \colhead{$N_{\rm diskbb}$} & \colhead{$\Gamma$} & \colhead{$\log\xi$} & \colhead{$R_{\rm refl}$} & \colhead{norm($\times10^{-2}$)} & \colhead{$\chi^2_{red}$}
}

\startdata
\multicolumn{12}{l}{\textbf{Stage I (2026-01-18 -- 2026-02-20)}} \\
\hline
P081404700102-20260118-01-01 & \texttt{relxill} & $1.35$ & $1.29$ & $1.43 \pm 0.08$ & -- & -- & $2.29 \pm 0.04$ & $0.3 \pm 4.0$ & $2.87 \pm 0.45$ & $2.03 \pm 0.09$ & $0.98$ \\
P081404700103-20260118-01-01 & \texttt{relxill} & $1.34$ & $1.33$ & $1.31 \pm 0.11$ & -- & -- & $2.37 \pm 0.03$ & $0.3 \pm 4.7$ & $4.39 \pm 0.57$ & $2.06 \pm 0.07$ & $0.97$ \\
P081404800102-20260126-01-01 & \texttt{relxill} & $1.33$ & $1.35$ & $0.37 \pm 0.10$ & -- & -- & $2.37 \pm 0.04$ & $3.79 \pm 0.17$ & $1.75 \pm 0.65$ & $1.97 \pm 0.58$ & $1.01$ \\
P081404800104-20260126-01-01 & \texttt{relxill} & $2.07$ & $1.81$ & $0.23 \pm 0.12$ & -- & -- & $2.47 \pm 0.05$ & $4.49 \pm 0.40$ & $2.15 \pm 0.90$ & $1.62 \pm 0.41$ & $1.00$ \\
P081404700201-20260127-01-01 & \texttt{relxill} & $1.10$ & $0.98$ & $0.45 \pm 0.14$ & -- & -- & $2.44 \pm 0.09$ & $3.89 \pm 0.37$ & $1.21 \pm 0.65$ & $3.2 \pm 1.4$ & $0.94$ \\
P081404700202-20260127-01-01 & \texttt{diskbb+powerlaw} & $1.57$ & $1.25$ & $0.19 \pm 0.10$ & $1.12 \pm 0.05$ & $86 \pm 12$ & $2.39 \pm 0.03$ & -- & -- & -- & $1.19$ \\
P081404700203-20260127-01-01 & \texttt{relxill} & $1.79$ & $2.24$ & $0.69 \pm 0.12$ & -- & -- & $2.67 \pm 0.07$ & $4.30 \pm 0.19$ & $3.2 \pm 2.1$ & $2.8 \pm 1.6$ & $1.06$ \\
P081404800301-20260128-01-01 & \texttt{diskbb+powerlaw} & $1.72$ & $1.43$ & $0.38 \pm 0.09$ & $1.15 \pm 0.02$ & $185 \pm 13$ & $2.48 \pm 0.03$ & -- & -- & -- & $1.09$ \\
P081404800302-20260128-01-01 & \texttt{relxill} & $1.32$ & $1.78$ & $1.23 \pm 0.08$ & -- & -- & $2.73 \pm 0.04$ & $4.70 \pm 0.26$ & $2.7 \pm 1.5$ & $5.8 \pm 2.8$ & $0.99$ \\
P081404800303-20260128-01-01 & \texttt{relxill} & $1.00$ & $1.20$ & $1.16 \pm 0.08$ & -- & -- & $2.63 \pm 0.07$ & $4.70 \pm 0.55$ & $1.10 \pm 0.84$ & $8.9 \pm 4.5$ & $0.93$ \\
P081404800601-20260129-01-01 & \texttt{diskbb+relxill} & $1.60$ & $1.55$ & $0.84 \pm 0.09$ & $1.15 \pm 0.02$ & $345 \pm 20$ & $2.57 \pm 0.04$ & $4.06 \pm 0.32$ & $1.50 \pm 0.62$ & $4.3 \pm 1.4$ & $0.92$ \\
P081404800602-20260129-01-01 & \texttt{relxill} & $1.18$ & $1.68$ & $2.26 \pm 0.06$ & -- & -- & $3.05 \pm 0.03$ & $4.70 \pm 0.53$ & $0.73 \pm 0.13$ & $50.9 \pm 6.2$ & $1.14$ \\
P081404800603-20260129-01-01 & \texttt{relxill} & $1.09$ & $1.25$ & $1.64 \pm 0.09$ & -- & -- & $2.81 \pm 0.05$ & $4.70 \pm 0.52$ & $1.23 \pm 0.53$ & $20.6 \pm 6.6$ & $1.07$ \\
P081404800701-20260130-01-01 & \texttt{relxill} & $1.09$ & $1.19$ & $0.81 \pm 0.08$ & -- & -- & $2.58 \pm 0.05$ & $4.70 \pm 0.30$ & $1.38 \pm 0.76$ & $5.4 \pm 2.1$ & $1.04$ \\
P081404800702-20260130-01-01 & \texttt{relxill} & $1.41$ & $1.75$ & $0.93 \pm 0.07$ & -- & -- & $2.69 \pm 0.05$ & $4.65 \pm 0.12$ & $2.7 \pm 1.5$ & $4.1 \pm 2.0$ & $1.10$ \\
P081404800703-20260130-01-01 & \texttt{relxill} & $1.19$ & $1.34$ & $0.91 \pm 0.10$ & -- & -- & $2.66 \pm 0.08$ & $4.62 \pm 0.23$ & $1.7 \pm 1.3$ & $5.9 \pm 3.5$ & $1.01$ \\
P081404800801-20260131-01-01 & \texttt{relxill} & $0.94$ & $0.83$ & $0.30 \pm 0.08$ & -- & -- & $2.42 \pm 0.03$ & $2.68 \pm 0.46$ & $0.50 \pm 0.13$ & $4.75 \pm 0.32$ & $0.98$ \\
P081404800802-20260131-01-01 & \texttt{relxill} & $1.35$ & $1.50$ & $0.31 \pm 0.10$ & -- & -- & $2.44 \pm 0.07$ & $4.34 \pm 0.14$ & $2.2 \pm 1.2$ & $1.83 \pm 0.88$ & $1.04$ \\
P081404800803-20260131-01-01 & \texttt{relxill} & $1.43$ & $1.18$ & $0.39 \pm 0.10$ & -- & -- & $2.47 \pm 0.05$ & $4^*$ & $1.6 \pm 1.0$ & $2.41 \pm 0.90$ & $0.95$ \\
P081404800901-20260201-01-01 & \texttt{relxill} & $1.45$ & $1.65$ & $0.46 \pm 0.08$ & -- & -- & $2.51 \pm 0.04$ & $4.34 \pm 0.07$ & $3.8 \pm 1.8$ & $1.45 \pm 0.64$ & $0.97$ \\
P081404800902-20260201-01-01 & \texttt{relxill} & $1.10$ & $1.19$ & $0.43 \pm 0.07$ & -- & -- & $2.55 \pm 0.03$ & $2.17 \pm 0.38$ & $0.49 \pm 0.16$ & $5.77 \pm 0.36$ & $1.00$ \\
P081404801001-20260202-01-01 & \texttt{relxill} & $1.53$ & $1.68$ & $0.36 \pm 0.07$ & -- & -- & $2.41 \pm 0.04$ & $4.17 \pm 0.04$ & $4.6 \pm 2.1$ & $0.95 \pm 0.41$ & $0.95$ \\
P081404801002-20260202-01-01 & \texttt{relxill} & $1.17$ & $1.10$ & $0.20 \pm 0.06$ & -- & -- & $2.41 \pm 0.04$ & $3.47 \pm 0.54$ & $0.33 \pm 0.12$ & $3.60 \pm 0.48$ & $0.99$ \\
P081404801101-20260203-01-01 & \texttt{relxill} & $1.10$ & $1.07$ & $0.37 \pm 0.06$ & -- & -- & $2.48 \pm 0.03$ & $2.38 \pm 0.29$ & $0.58 \pm 0.13$ & $4.40 \pm 0.24$ & $0.97$ \\
P081404801102-20260204-02-01 & \texttt{relxill} & $1.31$ & $1.48$ & $0.22 \pm 0.09$ & -- & -- & $2.42 \pm 0.06$ & $4.31 \pm 0.13$ & $1.55 \pm 0.78$ & $1.96 \pm 0.74$ & $0.98$ \\
P081404801103-20260204-02-01 & \texttt{relxill} & $1.36$ & $1.72$ & $0.21 \pm 0.15$ & -- & -- & $2.43 \pm 0.10$ & $4.34 \pm 0.12$ & $4.3 \pm 3.9$ & $1.07 \pm 0.95$ & $1.03$ \\
P081404801301-20260206-01-01 & \texttt{relxill} & $1.37$ & $1.71$ & $0.44 \pm 0.11$ & -- & -- & $2.51 \pm 0.06$ & $4.63 \pm 0.06$ & $6.9 \pm 6.5$ & $0.91 \pm 0.84$ & $1.04$ \\
P081404801302-20260206-01-01 & \texttt{relxill} & $1.38$ & $1.52$ & $0.61 \pm 0.11$ & -- & -- & $2.63 \pm 0.06$ & $4.37 \pm 0.24$ & $1^*$ & $4.74 \pm 0.81$ & $1.02$ \\
P081404801401-20260207-01-01 & \texttt{relxill} & $1.24$ & $1.46$ & $0.32 \pm 0.12$ & -- & -- & $2.50 \pm 0.06$ & $4.29 \pm 0.15$ & $1.7 \pm 1.0$ & $2.5 \pm 1.1$ & $1.09$ \\
P081404801402-20260207-01-01 & \texttt{relxill} & $1.51$ & $2.08$ & $0.60 \pm 0.10$ & -- & -- & $2.58 \pm 0.04$ & $4.50 \pm 0.09$ & $4.2 \pm 3.1$ & $1.6 \pm 1.0$ & $1.05$ \\
P081404801403-20260207-01-01 & \texttt{relxill} & $1.89$ & $3.14$ & $0.67 \pm 0.11$ & -- & -- & $2.70 \pm 0.06$ & $4.30 \pm 0.16$ & $3.5 \pm 1.9$ & $2.3 \pm 1.1$ & $1.02$ \\
P081404801901-20260208-01-01 & \texttt{relxill} & $1.47$ & $2.16$ & $0.73 \pm 0.07$ & -- & -- & $2.65 \pm 0.06$ & $4.57 \pm 0.13$ & $2.8 \pm 1.7$ & $2.7 \pm 1.5$ & $1.10$ \\
P081404801903-20260208-01-01 & \texttt{relxill} & $1.64$ & $2.63$ & $1.01 \pm 0.07$ & -- & -- & $2.77 \pm 0.03$ & $4.70 \pm 0.24$ & $2.9 \pm 1.5$ & $3.9 \pm 1.7$ & $1.06$ \\
P081404802001-20260209-01-01 & \texttt{relxill} & $1.59$ & $2.45$ & $0.96 \pm 0.06$ & -- & -- & $2.75 \pm 0.03$ & $4.65 \pm 0.11$ & $2.7 \pm 1.3$ & $3.6 \pm 1.4$ & $1.05$ \\
P081404802002-20260209-01-01 & \texttt{diskbb+relxill} & $1.69$ & $1.83$ & $0.42 \pm 0.10$ & $1.03 \pm 0.04$ & $172 \pm 21$ & $2.49 \pm 0.05$ & $3.99 \pm 0.23$ & $2.04 \pm 0.97$ & $1.65 \pm 0.60$ & $1.02$ \\
P081404802003-20260209-01-01 & \texttt{relxill} & $1.44$ & $2.18$ & $0.94 \pm 0.09$ & -- & -- & $2.76 \pm 0.05$ & $4.70 \pm 0.64$ & $0.89 \pm 0.57$ & $6.6 \pm 2.5$ & $1.05$ \\
P081404801501-20260210-01-01 & \texttt{relxill} & $1.50$ & $3.22$ & $0.74 \pm 0.08$ & -- & -- & $2.66 \pm 0.07$ & $4.62 \pm 0.13$ & $3.7 \pm 3.3$ & $2.2 \pm 1.8$ & $1.10$ \\
P081404801502-20260210-01-01 & \texttt{relxill} & $1.67$ & $2.63$ & $1.00 \pm 0.05$ & -- & -- & $2.77 \pm 0.03$ & $4.70 \pm 0.18$ & $3.1 \pm 1.3$ & $3.5 \pm 1.2$ & $1.15$ \\
P081404801503-20260210-01-01 & \texttt{relxill} & $1.93$ & $3.65$ & $1.11 \pm 0.09$ & -- & -- & $2.81 \pm 0.05$ & $4.70 \pm 0.19$ & $8.6 \pm 6.7$ & $1.8 \pm 1.4$ & $1.13$ \\
P081404802102-20260212-01-01 & \texttt{relxill} & $1.59$ & $2.48$ & $0.91 \pm 0.06$ & -- & -- & $2.75 \pm 0.03$ & $4.70 \pm 0.24$ & $2.10 \pm 0.90$ & $4.0 \pm 1.4$ & $1.16$ \\
P081404802103-20260212-01-01 & \texttt{relxill} & $1.45$ & $2.31$ & $0.67 \pm 0.07$ & -- & -- & $2.66 \pm 0.05$ & $4.58 \pm 0.14$ & $2.3 \pm 1.2$ & $2.8 \pm 1.3$ & $1.22$ \\
P081404802301-20260215-01-01 & \texttt{diskbb+powerlaw} & $1.44$ & $1.21$ & $0.39 \pm 0.07$ & $1.06 \pm 0.01$ & $398 \pm 21$ & $2.48 \pm 0.03$ & -- & -- & -- & $0.98$ \\
P081404802302-20260215-01-01 & \texttt{diskbb+relxill} & $1.31$ & $1.61$ & $0.74 \pm 0.11$ & $0.99 \pm 0.03$ & $522 \pm 66$ & $2.56 \pm 0.06$ & $4.0 \pm 2.5$ & $0.18 \pm 0.30$ & $4.2 \pm 1.4$ & $0.99$ \\
P081404802501-20260217-01-01 & \texttt{diskbb+powerlaw} & $1.33$ & $1.32$ & $0.13 \pm 0.06$ & $1.05 \pm 0.02$ & $184 \pm 12$ & $2.40 \pm 0.02$ & -- & -- & -- & $1.00$ \\
P081404802502-20260217-01-01 & \texttt{diskbb+relxill} & $1.49$ & $1.57$ & $0.24 \pm 0.08$ & $1.02 \pm 0.02$ & $221 \pm 20$ & $2.43 \pm 0.05$ & $3.98 \pm 0.43$ & $0.69 \pm 0.40$ & $1.89 \pm 0.54$ & $0.91$ \\
P081404802602-20260219-01-01 & \texttt{relxill} & $1.77$ & $3.14$ & $0.82 \pm 0.06$ & -- & -- & $2.76 \pm 0.03$ & $4.70 \pm 0.18$ & $4.5 \pm 2.7$ & $1.9 \pm 1.0$ & $1.13$ \\
P081404802603-20260219-01-01 & \texttt{relxill} & $1.44$ & $2.57$ & $0.52 \pm 0.09$ & -- & -- & $2.70 \pm 0.06$ & $4.70 \pm 0.34$ & $2.5 \pm 2.0$ & $2.6 \pm 1.7$ & $1.05$ \\
\multicolumn{12}{l}{\textbf{Stage II (2026-02-21 -- 2026-04-01)}} \\
\hline
P081404802701-20260221-01-01 & \texttt{diskbb+powerlaw} & $1.50$ & $1.49$ & $0.31 \pm 0.05$ & $1.04 \pm 0.01$ & $369 \pm 15$ & $2.45 \pm 0.02$ & -- & -- & -- & $1.11$ \\
P081404802702-20260221-01-01 & \texttt{diskbb+powerlaw} & $1.36$ & $1.41$ & $0.40 \pm 0.07$ & $1.04 \pm 0.02$ & $311 \pm 18$ & $2.52 \pm 0.03$ & -- & -- & -- & $0.93$ \\
P081404802802-20260224-02-01 & \texttt{diskbb+relxill} & $1.50$ & $1.65$ & $0.10 \pm 0.12$ & $0.99 \pm 0.03$ & $205 \pm 27$ & $2.35 \pm 0.08$ & $4.10 \pm 0.17$ & $2.1 \pm 1.8$ & $0.76 \pm 0.51$ & $0.90$ \\
P081404802803-20260224-02-01 & \texttt{diskbb+powerlaw} & $1.26$ & $1.25$ & $0.15 \pm 0.11$ & $0.95 \pm 0.03$ & $267 \pm 40$ & $2.38 \pm 0.03$ & -- & -- & -- & $0.99$ \\
P081404802901-20260225-01-01 & \texttt{diskbb+powerlaw} & $1.33$ & $1.56$ & $0.34 \pm 0.12$ & $1.03 \pm 0.03$ & $271 \pm 32$ & $2.48 \pm 0.04$ & -- & -- & -- & $0.97$ \\
P081404802903-20260225-01-01 & \texttt{relxill} & $1.39$ & $3.70$ & $1.47 \pm 0.12$ & -- & -- & $3.20 \pm 0.05$ & $4.70 \pm 0.68$ & $1.66 \pm 0.51$ & $22.4 \pm 5.0$ & $1.28$ \\
P081404803002-20260226-01-01 & \texttt{diskbb+relxill} & $1.39$ & $1.71$ & $0.50 \pm 0.12$ & $0.93 \pm 0.03$ & $355 \pm 59$ & $2.45 \pm 0.08$ & $3.95 \pm 0.37$ & $1.4 \pm 1.1$ & $1.35 \pm 0.68$ & $0.95$ \\
P081404803102-20260227-01-01 & \texttt{diskbb+powerlaw} & $1.23$ & $1.13$ & $0.15 \pm 0.12$ & $0.90 \pm 0.04$ & $253 \pm 47$ & $2.30 \pm 0.03$ & -- & -- & -- & $1.04$ \\
P081404803203-20260301-02-01 & \texttt{diskbb+powerlaw} & $1.48$ & $1.49$ & $0.10 \pm 0.08$ & $1.04 \pm 0.02$ & $240 \pm 18$ & $2.33 \pm 0.03$ & -- & -- & -- & $0.94$ \\
P081404803302-20260301-01-01 & \texttt{diskbb+relxill} & $1.60$ & $1.80$ & $0.19 \pm 0.12$ & $1.05 \pm 0.02$ & $227 \pm 18$ & $2.56 \pm 0.09$ & $2.3 \pm 1.0$ & $0.82 \pm 0.37$ & $2.13 \pm 0.66$ & $0.92$ \\
P081404803701-20260303-01-01 & \texttt{diskbb+relxill} & $1.23$ & $2.08$ & $0.49 \pm 0.17$ & $0.91 \pm 0.06$ & $360 \pm 100$ & $2.48 \pm 0.07$ & $4.0 \pm 2.0$ & $0.30 \pm 0.72$ & $2.2 \pm 1.2$ & $0.87$ \\
P081404803702-20260303-01-01 & \texttt{diskbb+powerlaw} & $1.23$ & $1.30$ & $0.26 \pm 0.07$ & $0.96 \pm 0.02$ & $307 \pm 22$ & $2.41 \pm 0.03$ & -- & -- & -- & $0.96$ \\
P081404803703-20260303-01-01 & \texttt{diskbb+relxill} & $1.54$ & $2.07$ & $0.27 \pm 0.13$ & $1.01 \pm 0.03$ & $248 \pm 30$ & $2.42 \pm 0.10$ & $4.00 \pm 0.95$ & $0.57 \pm 0.80$ & $1.34 \pm 0.75$ & $0.90$ \\
P081404803401-20260304-01-01 & \texttt{diskbb+relxill} & $1.09$ & $1.15$ & $0.36 \pm 0.21$ & $0.91 \pm 0.03$ & $365 \pm 49$ & $2.19 \pm 0.22$ & $4.7 \pm 1.4$ & $0.9 \pm 1.8$ & $0.9 \pm 1.1$ & $0.90$ \\
P081404803801-20260306-01-01 & \texttt{diskbb+powerlaw} & $1.03$ & $1.15$ & $0.20 \pm 0.09$ & $0.91 \pm 0.02$ & $361 \pm 41$ & $2.41 \pm 0.03$ & -- & -- & -- & $1.02$ \\
P081404803802-20260307-02-01 & \texttt{diskbb+powerlaw} & $1.34$ & $1.40$ & $0.13 \pm 0.07$ & $0.95 \pm 0.02$ & $311 \pm 22$ & $2.40 \pm 0.04$ & -- & -- & -- & $0.99$ \\
P081404804101-20260309-01-01 & \texttt{relxill} & $1.15$ & $4.13$ & $1.58 \pm 0.08$ & -- & -- & $3.40 \pm 0.10$ & $0.00$ & $0.82 \pm 0.59$ & $43.7 \pm 6.3$ & $1.22$ \\
P081404804501-20260313-01-01 & \texttt{diskbb+powerlaw} & $4.52$ & $5.12$ & $0.14 \pm 0.06$ & $0.83 \pm 0.01$ & $645 \pm 32$ & $1.76 \pm 0.11$ & -- & -- & -- & $0.99$ \\
P081404804701-20260316-01-01 & \texttt{diskbb+powerlaw} & $0.67$ & $0.34$ & $0.27 \pm 0.12$ & $0.77 \pm 0.02$ & $790 \pm 100$ & $1.80 \pm 0.25$ & -- & -- & -- & $0.90$ \\
P081404801601-20260318-01-01 & \texttt{diskbb+powerlaw} & $3.50$ & $6.50$ & $0.22 \pm 0.09$ & $0.79 \pm 0.01$ & $670 \pm 58$ & $2.08 \pm 0.13$ & -- & -- & -- & $0.93$ \\
P081404804901-20260320-01-01 & \texttt{diskbb+powerlaw} & $16.81$ & $32.51$ & $0.68 \pm 0.13$ & $0.71 \pm 0.01$ & $1180 \pm 130$ & $2.12 \pm 0.18$ & -- & -- & -- & $1.35$ \\
P081404805001-20260322-01-01 & \texttt{diskbb+powerlaw} & $20.22$ & $26.00$ & $0.63 \pm 0.14$ & $0.69 \pm 0.01$ & $1210 \pm 140$ & $1.70 \pm 0.24$ & -- & -- & -- & $1.30$ \\
P081404805101-20260323-01-01 & \texttt{diskbb+powerlaw} & $22.39$ & $30.88$ & $1.32 \pm 0.28$ & $0.65 \pm 0.02$ & $1990 \pm 460$ & $1.83 \pm 0.29$ & -- & -- & -- & $1.08$ \\
\enddata

\tablecomments{$N_{\rm H}$ is given in units of $10^{22}~{\rm cm}^{-2}$. For relxill-based models, $A_{\rm Fe}$ was fixed at 1. The superscript $*$ marks fixed parameters.}
\end{deluxetable*}

\onecolumngrid
\begin{center}
\makeatletter\def\table@hook{\scriptsize}\makeatother 
\scriptsize
\setlength{\tabcolsep}{4.5pt}
\begin{longtable}{l c | ccc | ccc | ccc}
\caption{Evolution of QPO Parameters (Frequency, Q-value, and RMS) across LE, ME, and HE} \label{tab:qpo_params_final} \\
\hline\hline
\multirow{2}{*}{Exposure ID} & \multirow{2}{*}{Avg MJD} & \multicolumn{3}{c|}{LE Band (2--10 keV)} & \multicolumn{3}{c|}{ME Band (10--30 keV)} & \multicolumn{3}{c}{HE Band (27--250 keV)} \\
\cline{3-11}
& & $\nu_{\rm QPO}$ (Hz) & $Q$-value & RMS (\%) & $\nu_{\rm QPO}$ (Hz) & $Q$-value & RMS (\%) & $\nu_{\rm QPO}$ (Hz) & $Q$-value & RMS (\%) \\
\hline
\endfirsthead
\multicolumn{11}{c}{\tablename\ \thetable\ -- \textit{Continued from previous page}} \\
\hline\hline
\multirow{2}{*}{Exposure ID} & \multirow{2}{*}{Avg MJD} & \multicolumn{3}{c|}{LE Band (2--10 keV)} & \multicolumn{3}{c|}{ME Band (10--30 keV)} & \multicolumn{3}{c}{HE Band (27--250 keV)} \\
\cline{3-11}
& & $\nu_{\rm QPO}$ (Hz) & $Q$-value & RMS (\%) & $\nu_{\rm QPO}$ (Hz) & $Q$-value & RMS (\%) & $\nu_{\rm QPO}$ (Hz) & $Q$-value & RMS (\%) \\
\hline
\endhead
\hline \multicolumn{11}{r}{\textit{Continued on next page}} \\
\endfoot
\hline
\multicolumn{11}{l}{\footnotesize \textit{Note:} For the Quality Factor ($Q$), errors were derived via standard error propagation from the centroid frequency and FWHM. In specific high-energy exposures, where} \\ \multicolumn{11}{l}{\footnotesize the QPO width was statistically unconstrained due to low signal-to-noise, the resulting unphysical error bounds on $Q$ have been omitted for clarity.} \\
\endlastfoot
P081404700101 & 61058.14 & -- & -- & -- & $0.26 \pm 0.01$ & $4.1 \pm 2.5$ & $16.8 \pm 2.9$ & $0.25 \pm 0.01$ & $9.9 \pm 6.0$ & $15.6 \pm 2.7$ \\
P081404700102 & 61058.31 & $0.30 \pm 0.01$ & $11.1 \pm 5.8$ & $10.0 \pm 1.7$ & $0.29 \pm 0.01$ & $5.6 \pm 1.9$ & $15.5 \pm 1.8$ & $0.30 \pm 0.01$ & $4.6 \pm 2.1$ & $19.5 \pm 3.0$ \\
P081404700103 & 61058.42 & $0.32 \pm 0.05$ & $2.0 \pm 2.1$ & $11.5 \pm 5.0$ & $0.29 \pm 0.02$ & $2.7 \pm 1.1$ & $18.9 \pm 2.9$ & $0.27 \pm 0.01$ & $3.3 \pm 1.1$ & $19.2 \pm 2.7$ \\
P081404700201 & 61067.13 & -- & -- & -- & $3.36 \pm 0.09$ & $4.6 \pm 1.3$ & $22.3 \pm 2.2$ & $3.48 \pm 0.21$ & $6.0 \pm 5.9$ & $19.1 \pm 5.3$ \\
P081404700202 & 61067.24 & $3.68 \pm 0.09$ & $4.4 \pm 1.1$ & $9.5 \pm 0.9$ & $3.52 \pm 0.11$ & $3.8 \pm 1.2$ & $23.2 \pm 2.4$ & $3.54 \pm 0.08$ & $9.1 \pm 4.7$ & $14.5 \pm 2.4$ \\
P081404700203 & 61067.38 & $3.97 \pm 0.05$ & $18.3 \pm 9.4$ & $6.8 \pm 1.1$ & $4.01 \pm 0.06$ & $26 \pm 24$ & $15.3 \pm 2.9$ & -- & -- & -- \\
P081404800101 & 61066.08 & -- & -- & -- & $2.32 \pm 0.03$ & $7.9 \pm 2.3$ & $21.5 \pm 2.2$ & $2.30 \pm 0.04$ & $13.0 \pm 8.3$ & $17.2 \pm 2.9$ \\
P081404800102 & 61066.21 & $2.65 \pm 0.04$ & $14.4 \pm 6.8$ & $6.6 \pm 0.9$ & $2.58 \pm 0.04$ & $5.9 \pm 1.5$ & $22.6 \pm 1.8$ & $2.59 \pm 0.04$ & $7.9 \pm 3.4$ & $16.6 \pm 2.0$ \\
P081404800103 & 61066.35 & $2.80 \pm 0.04$ & $11.8 \pm 4.0$ & $8.3 \pm 0.9$ & $2.73 \pm 0.03$ & $10.2 \pm 2.8$ & $20.9 \pm 1.8$ & $2.76 \pm 0.03$ & $14.3 \pm 5.8$ & $17.1 \pm 1.9$ \\
P081404800104 & 61066.50 & $2.82 \pm 0.05$ & $9.0 \pm 3.5$ & $9.6 \pm 1.1$ & $2.77 \pm 0.03$ & $8.9 \pm 2.7$ & $20.4 \pm 2.0$ & $2.85 \pm 0.05$ & $13.1 \pm 7.6$ & $20.9 \pm 3.4$ \\
P081404800301 & 61068.32 & $5.69 \pm 0.10$ & $19 \pm 20$ & $3.9 \pm 0.5$ & $6.02 \pm 0.24$ & $14 \pm 25$ & $9.2 \pm 4.4$ & -- & -- & -- \\
P081404800302 & 61068.46 & $5.82 \pm 0.20$ & $11.8 \pm 9.8$ & $4.7 \pm 1.2$ & -- & -- & -- & -- & -- & -- \\
P081404800701 & 61070.52 & -- & -- & -- & $6.30 \pm 0.21$ & $6.6 \pm 4.1$ & $15.0 \pm 3.2$ & -- & -- & -- \\
P081404800702 & 61070.65 & $5.32 \pm 0.22$ & $5.5 \pm 2.8$ & $6.0 \pm 1.2$ & $5.06 \pm 0.28$ & $4.4 \pm 1.9$ & $16.9 \pm 2.6$ & -- & -- & -- \\
P081404800801 & 61071.51 & $3.62 \pm 0.05$ & $11.9 \pm 4.8$ & $7.6 \pm 1.1$ & $3.67 \pm 0.05$ & $8.4 \pm 2.4$ & $20.7 \pm 2.1$ & $3.76 \pm 0.14$ & $5.2 \pm 2.8$ & $18.0 \pm 3.1$ \\
P081404800802 & 61071.64 & $3.53 \pm 0.03$ & $13.5 \pm 4.6$ & $8.4 \pm 0.8$ & $3.53 \pm 0.04$ & $11.3 \pm 3.4$ & $18.4 \pm 2.0$ & $3.43 \pm 0.06$ & $15 \pm 12$ & $12.8 \pm 2.8$ \\
P081404800803 & 61071.77 & $3.22 \pm 0.06$ & $16 \pm 16$ & $7.0 \pm 1.6$ & $3.26 \pm 0.10$ & $7.7 \pm 4.7$ & $17.4 \pm 3.3$ & $3.26 \pm 0.05$ & $11.3 \pm 6.1$ & $21.8 \pm 3.5$ \\
P081404800901 & 61072.63 & $3.67 \pm 0.03$ & $10.7 \pm 3.3$ & $7.5 \pm 0.6$ & $3.61 \pm 0.05$ & $8.2 \pm 3.5$ & $20.8 \pm 2.3$ & $3.69 \pm 0.06$ & $8.3 \pm 3.0$ & $17.7 \pm 2.0$ \\
P081404800902 & 61072.83 & $3.73 \pm 0.06$ & $9.0 \pm 3.2$ & $8.0 \pm 0.9$ & $3.78 \pm 0.08$ & $7.0 \pm 2.5$ & $20.2 \pm 2.5$ & $3.76 \pm 0.13$ & $6.5 \pm 3.9$ & $14.6 \pm 3.0$ \\
P081404801001 & 61073.68 & $3.13 \pm 0.05$ & $6.8 \pm 2.0$ & $8.6 \pm 0.9$ & $3.15 \pm 0.07$ & $5.7 \pm 1.5$ & $20.5 \pm 2.0$ & $3.15 \pm 0.08$ & $6.3 \pm 2.4$ & $17.4 \pm 2.2$ \\
P081404801002 & 61073.88 & $3.24 \pm 0.05$ & $7.6 \pm 2.2$ & $8.8 \pm 1.0$ & $3.28 \pm 0.05$ & $7.4 \pm 2.4$ & $22.0 \pm 2.1$ & $3.17 \pm 0.06$ & - & $15.9 \pm 4.5$ \\
P081404801101 & 61074.87 & $3.38 \pm 0.03$ & $11.2 \pm 4.0$ & $8.0 \pm 0.9$ & $3.40 \pm 0.04$ & $9.9 \pm 4.3$ & $18.3 \pm 2.5$ & $3.40 \pm 0.20$ & - & $11.3 \pm 5.7$ \\
P081404801102 & 61075.07 & $3.66 \pm 0.08$ & $9.4 \pm 4.9$ & $6.3 \pm 1.1$ & $3.58 \pm 0.07$ & $9.1 \pm 5.7$ & $17.3 \pm 2.9$ & $3.66 \pm 0.06$ & $14 \pm 12$ & $13.6 \pm 2.9$ \\
P081404801103 & 61075.20 & $3.80 \pm 0.09$ & $9.0 \pm 8.0$ & $7.5 \pm 1.4$ & $4.00 \pm 0.10$ & $8.2 \pm 4.4$ & $21.4 \pm 3.7$ & $3.89 \pm 0.36$ & $9 \pm 12$ & $12.1 \pm 4.5$ \\
P081404801301 & 61077.04 & $4.26 \pm 0.08$ & $14 \pm 13$ & $5.9 \pm 1.2$ & $4.73 \pm 0.24$ & $5.5 \pm 5.5$ & $17.9 \pm 4.9$ & $4.64 \pm 0.35$ & $6.3 \pm 5.6$ & $14.2 \pm 4.2$ \\
P081404801302 & 61077.18 & $4.92 \pm 0.08$ & $15 \pm 11$ & $6.7 \pm 1.1$ & $4.43 \pm 0.31$ & $4.9 \pm 4.6$ & $18.7 \pm 5.1$ & -- & -- & -- \\
P081404801401 & 61078.36 & $4.68 \pm 0.13$ & $6.6 \pm 3.4$ & $7.8 \pm 1.2$ & $4.30 \pm 0.17$ & $6.1 \pm 3.1$ & $20.0 \pm 3.6$ & $4.71 \pm 0.19$ & $19 \pm 23$ & $13.5 \pm 4.8$ \\
P081404801402 & 61078.43 & $3.99 \pm 0.07$ & $13.7 \pm 8.4$ & $5.2 \pm 1.0$ & $4.31 \pm 0.33$ & $2.5 \pm 1.4$ & $25.1 \pm 4.3$ & -- & -- & -- \\
P081404801403 & 61078.55 & $3.85 \pm 0.14$ & - & $3.7 \pm 0.8$ & $3.88 \pm 0.07$ & $18 \pm 17$ & $18.9 \pm 4.5$ & $3.76 \pm 0.17$ & $8.2 \pm 7.5$ & $13.8 \pm 3.8$ \\
P081404801501 & 61081.18 & $4.85 \pm 0.11$ & $12.4 \pm 9.7$ & $4.7 \pm 1.1$ & -- & -- & -- & -- & -- & -- \\
P081404801502 & 61081.32 & $4.88 \pm 0.15$ & $5.8 \pm 2.7$ & $5.4 \pm 1.0$ & $4.93 \pm 0.15$ & $11 \pm 13$ & $12.9 \pm 4.0$ & -- & -- & -- \\
P081404801901 & 61079.06 & $4.65 \pm 0.06$ & $22 \pm 20$ & $4.5 \pm 1.0$ & $4.72 \pm 0.17$ & $5.3 \pm 3.0$ & $20.1 \pm 3.6$ & -- & -- & -- \\
P081404801903 & 61079.33 & $5.10 \pm 0.05$ & $19 \pm 15$ & $5.0 \pm 0.9$ & -- & -- & -- & -- & -- & -- \\
P081404802001 & 61080.39 & $4.83 \pm 0.11$ & $8.1 \pm 4.5$ & $5.4 \pm 0.9$ & $4.83 \pm 0.19$ & $6.8 \pm 5.0$ & $17.4 \pm 3.9$ & -- & -- & -- \\
P081404802002 & 61080.54 & $5.44 \pm 0.07$ & $17 \pm 10$ & $4.4 \pm 0.8$ & $5.68 \pm 0.15$ & $12 \pm 14$ & $14.2 \pm 4.4$ & -- & -- & -- \\
P081404802003 & 61080.66 & $5.11 \pm 0.09$ & $17 \pm 14$ & $5.0 \pm 1.1$ & $4.90 \pm 0.22$ & $13 \pm 14$ & $15.6 \pm 5.6$ & -- & -- & -- \\
P081404802101 & 61083.08 & $4.74 \pm 0.10$ & $8.4 \pm 3.6$ & $6.3 \pm 1.0$ & $4.64 \pm 0.20$ & $6.9 \pm 4.9$ & $15.7 \pm 4.1$ & $4.89 \pm 0.36$ & $5.3 \pm 4.1$ & $21.2 \pm 5.7$ \\
P081404802102 & 61083.23 & $5.22 \pm 0.16$ & $8.8 \pm 8.4$ & $4.5 \pm 1.2$ & $5.15 \pm 0.12$ & $8.8 \pm 5.6$ & $17.3 \pm 3.4$ & -- & -- & -- \\
P081404802103 & 61083.39 & $4.90 \pm 0.09$ & $9.9 \pm 5.8$ & $5.3 \pm 0.9$ & $5.05 \pm 0.15$ & $5.9 \pm 2.9$ & $19.0 \pm 3.1$ & -- & -- & -- \\
P081404802201 & 61085.09 & -- & -- & -- & $6.09 \pm 0.39$ & $2.6 \pm 1.2$ & $25.8 \pm 4.0$ & -- & -- & -- \\
P081404802601 & 61090.48 & $5.13 \pm 0.11$ & $17 \pm 18$ & $3.9 \pm 1.1$ & -- & -- & -- & -- & -- & -- \\
P081404802602 & 61090.63 & $5.05 \pm 0.07$ & - & $3.4 \pm 1.3$ & -- & -- & -- & -- & -- & -- \\
P081404802603 & 61090.75 & $5.44 \pm 0.48$ & - & $3.9 \pm 2.5$ & -- & -- & -- & -- & -- & -- \\
P081404803101 & 61098.89 & $6.52 \pm 0.36$ & $9.3 \pm 9.9$ & $4.4 \pm 1.5$ & -- & -- & -- & -- & -- & -- \\
P081404803301 & 61100.70 & -- & -- & -- & $4.94 \pm 0.15$ & $13 \pm 10$ & $15.7 \pm 4.1$ & -- & -- & -- \\
\end{longtable}
\normalsize
\end{center}

\begin{center}
\makeatletter\def\table@hook{\scriptsize}\makeatother 
\scriptsize
\setlength{\tabcolsep}{4pt}
\begin{longtable}{l c c | rr | rr | rr}
\caption{Broadband Time and Phase Lags Relative to the 2--4 keV Reference Band.} \label{tab:lag_summary} \\
\hline\hline
\multirow{2}{*}{Exposure ID} & \multirow{2}{*}{MJD} & \multirow{2}{*}{$\nu_{\rm QPO}$ (Hz)} & \multicolumn{2}{c|}{LE (4--10 keV)} & \multicolumn{2}{c|}{ME (10--30 keV)} & \multicolumn{2}{c}{HE (27--250 keV)} \\
\cline{4-9}
& & & Time Lag (s) & Phase (rad) & Time Lag (s) & Phase (rad) & Time Lag (s) & Phase (rad) \\
\hline
\endfirsthead
\multicolumn{9}{c}{\tablename\ \thetable\ -- \textit{Continued from previous page}} \\
\hline\hline
\multirow{2}{*}{Exposure ID} & \multirow{2}{*}{MJD} & \multirow{2}{*}{$\nu_{\rm QPO}$ (Hz)} & \multicolumn{2}{c|}{LE (4--10 keV)} & \multicolumn{2}{c|}{ME (10--30 keV)} & \multicolumn{2}{c}{HE (27--250 keV)} \\
\cline{4-9}
& & & Time Lag (s) & Phase (rad) & Time Lag (s) & Phase (rad) & Time Lag (s) & Phase (rad) \\
\hline
\endhead
\hline \multicolumn{9}{r}{\textit{Continued on next page}} \\
\endfoot
\hline
\endlastfoot
P081404700102 & 61058.31 & 0.29 & $-0.012 \pm 0.061$ & $-0.02 \pm 0.11$ & $-0.057 \pm 0.060$ & $-0.10 \pm 0.11$ & $-0.083 \pm 0.064$ & $-0.15 \pm 0.11$ \\
P081404700103 & 61058.42 & 0.29 & $0.046 \pm 0.078$ & $0.08 \pm 0.14$ & $0.16 \pm 0.11$ & $0.29 \pm 0.19$ & $0.06 \pm 0.10$ & $0.10 \pm 0.18$ \\
P081404700201 & 61067.13 & 3.42 & $-0.027 \pm 0.046$ & $-0.6 \pm 1.0$ & -- & -- & $-0.01 \pm 0.10$ & $-0.1 \pm 2.3$ \\
P081404700202 & 61067.25 & 3.58 & $-0.005 \pm 0.009$ & $-0.11 \pm 0.19$ & $0.024 \pm 0.045$ & $0.54 \pm 0.99$ & $0.022 \pm 0.041$ & $0.49 \pm 0.89$ \\
P081404700203 & 61067.39 & 3.98 & $0.001 \pm 0.007$ & $0.03 \pm 0.17$ & $0.004 \pm 0.012$ & $0.09 \pm 0.29$ & -- & -- \\
P081404800102 & 61066.19 & 2.60 & -- & -- & $0.117 \pm 0.071$ & $2.0 \pm 1.2$ & $0.090 \pm 0.030$ & $1.49 \pm 0.50$ \\
P081404800103 & 61066.32 & 2.76 & $-0.004 \pm 0.007$ & $-0.06 \pm 0.13$ & $0.015 \pm 0.008$ & $0.27 \pm 0.13$ & $-0.002 \pm 0.011$ & $-0.03 \pm 0.20$ \\
P081404800104 & 61066.49 & 2.81 & $0.019 \pm 0.010$ & $0.34 \pm 0.17$ & $0.007 \pm 0.011$ & $0.13 \pm 0.19$ & -- & -- \\
P081404800301 & 61068.31 & 5.86 & $-0.001 \pm 0.015$ & $-0.05 \pm 0.55$ & -- & -- & $0.038 \pm 0.022$ & $1.42 \pm 0.82$ \\
P081404800302 & 61068.46 & 5.82 & -- & -- & $0.036 \pm 0.016$ & $1.28 \pm 0.56$ & -- & -- \\
P081404800701 & 61070.46 & 6.30 & $-0.014 \pm 0.012$ & $-0.56 \pm 0.50$ & -- & -- & -- & -- \\
P081404800702 & 61070.64 & 5.19 & $-0.020 \pm 0.007$ & $-0.63 \pm 0.24$ & -- & -- & $0.011 \pm 0.016$ & $0.36 \pm 0.51$ \\
P081404800801 & 61071.47 & 3.68 & $0.019 \pm 0.006$ & $0.45 \pm 0.14$ & $0.007 \pm 0.008$ & $0.17 \pm 0.19$ & $0.018 \pm 0.007$ & $0.43 \pm 0.17$ \\
P081404800802 & 61071.63 & 3.50 & $0.003 \pm 0.006$ & $0.06 \pm 0.13$ & $0.003 \pm 0.006$ & $0.06 \pm 0.14$ & $0.022 \pm 0.013$ & $0.48 \pm 0.29$ \\
P081404800803 & 61071.76 & 3.24 & $-0.011 \pm 0.012$ & $-0.24 \pm 0.25$ & $0.004 \pm 0.012$ & $0.08 \pm 0.26$ & $0.019 \pm 0.017$ & $0.40 \pm 0.35$ \\
P081404800901 & 61072.63 & 3.65 & $0.004 \pm 0.004$ & $0.09 \pm 0.10$ & $0.003 \pm 0.006$ & $0.06 \pm 0.13$ & $0.008 \pm 0.009$ & $0.19 \pm 0.22$ \\
P081404800902 & 61072.83 & 3.76 & $0.004 \pm 0.005$ & $0.10 \pm 0.12$ & $0.007 \pm 0.007$ & $0.17 \pm 0.16$ & $-0.019 \pm 0.012$ & $-0.43 \pm 0.28$ \\
P081404801001 & 61073.68 & 3.14 & $0.006 \pm 0.006$ & $0.12 \pm 0.11$ & $0.012 \pm 0.008$ & $0.24 \pm 0.15$ & $0.004 \pm 0.011$ & $0.08 \pm 0.22$ \\
P081404801002 & 61073.88 & 3.23 & $0.009 \pm 0.006$ & $0.18 \pm 0.13$ & $0.010 \pm 0.006$ & $0.21 \pm 0.13$ & $0.019 \pm 0.011$ & $0.40 \pm 0.23$ \\
P081404801101 & 61074.87 & 3.40 & $-0.003 \pm 0.004$ & $-0.061 \pm 0.091$ & $0.010 \pm 0.006$ & $0.20 \pm 0.12$ & $0.033 \pm 0.016$ & $0.69 \pm 0.33$ \\
P081404801102 & 61075.07 & 3.63 & $-0.001 \pm 0.009$ & $-0.03 \pm 0.20$ & $-0.008 \pm 0.012$ & $-0.18 \pm 0.27$ & $0.020 \pm 0.010$ & $0.47 \pm 0.24$ \\
P081404801103 & 61075.18 & 3.90 & $0.006 \pm 0.011$ & $0.15 \pm 0.28$ & $0.017 \pm 0.022$ & $0.42 \pm 0.55$ & -- & -- \\
P081404801301 & 61077.04 & 4.54 & $0.007 \pm 0.008$ & $0.21 \pm 0.23$ & $0.002 \pm 0.011$ & $0.06 \pm 0.31$ & -- & -- \\
P081404801401 & 61078.31 & 4.56 & $-0.014 \pm 0.058$ & $-0.4 \pm 1.7$ & $-0.015 \pm 0.015$ & $-0.43 \pm 0.43$ & $0.022 \pm 0.015$ & $0.64 \pm 0.45$ \\
P081404801402 & 61078.42 & 4.15 & $0.003 \pm 0.011$ & $0.08 \pm 0.29$ & $0.016 \pm 0.015$ & $0.42 \pm 0.38$ & $-0.000 \pm 0.023$ & $0.00 \pm 0.58$ \\
P081404801403 & 61078.55 & 3.83 & $0.004 \pm 0.015$ & $0.11 \pm 0.37$ & $-0.001 \pm 0.009$ & $-0.02 \pm 0.22$ & -- & -- \\
P081404801501 & 61081.13 & 4.85 & $0.012 \pm 0.013$ & $0.35 \pm 0.41$ & $0.012 \pm 0.008$ & $0.37 \pm 0.25$ & -- & -- \\
P081404801502 & 61081.26 & 4.90 & $-0.009 \pm 0.012$ & $-0.29 \pm 0.37$ & $0.017 \pm 0.013$ & $0.52 \pm 0.39$ & $0.007 \pm 0.014$ & $0.20 \pm 0.43$ \\
P081404801901 & 61079.02 & 4.68 & $-0.007 \pm 0.007$ & $-0.20 \pm 0.20$ & $0.024 \pm 0.009$ & $0.69 \pm 0.25$ & $-0.015 \pm 0.017$ & $-0.43 \pm 0.48$ \\
P081404801903 & 61079.29 & 5.10 & $0.006 \pm 0.008$ & $0.18 \pm 0.24$ & $0.002 \pm 0.014$ & $0.08 \pm 0.45$ & $0.022 \pm 0.020$ & $0.71 \pm 0.65$ \\
P081404802001 & 61080.34 & 4.83 & $0.010 \pm 0.011$ & $0.30 \pm 0.33$ & $-0.004 \pm 0.013$ & $-0.13 \pm 0.41$ & $-0.003 \pm 0.011$ & $-0.10 \pm 0.34$ \\
P081404802002 & 61080.52 & 5.56 & $0.007 \pm 0.007$ & $0.24 \pm 0.25$ & $-0.015 \pm 0.015$ & $-0.54 \pm 0.53$ & $0.004 \pm 0.024$ & $0.14 \pm 0.84$ \\
P081404802003 & 61080.66 & 5.01 & $0.011 \pm 0.008$ & $0.36 \pm 0.25$ & $0.020 \pm 0.017$ & $0.64 \pm 0.55$ & $0.032 \pm 0.033$ & $1.0 \pm 1.1$ \\
P081404802101 & 61083.04 & 4.76 & $-0.029 \pm 0.014$ & $-0.87 \pm 0.42$ & $0.091 \pm 0.020$ & $2.78 \pm 0.61$ & -- & -- \\
P081404802102 & 61083.17 & 5.18 & $-0.001 \pm 0.009$ & $-0.03 \pm 0.30$ & $0.00 \pm 0.14$ & $-0.1 \pm 4.3$ & $-0.002 \pm 0.028$ & $-0.05 \pm 0.91$ \\
P081404802103 & 61083.35 & 4.98 & $-0.002 \pm 0.006$ & $-0.07 \pm 0.19$ & $0.024 \pm 0.016$ & $0.74 \pm 0.50$ & -- & -- \\
P081404802201 & 61085.01 & 6.09 & -- & -- & -- & -- & $-0.061 \pm 0.016$ & $-2.29 \pm 0.60$ \\
P081404802601 & 61090.41 & 5.13 & $0.004 \pm 0.013$ & $0.13 \pm 0.41$ & $0.051 \pm 0.096$ & $1.7 \pm 3.1$ & $-0.021 \pm 0.016$ & $-0.67 \pm 0.50$ \\
P081404802602 & 61090.60 & 5.05 & $0.012 \pm 0.010$ & $0.40 \pm 0.33$ & $0.026 \pm 0.010$ & $0.83 \pm 0.33$ & $0.015 \pm 0.040$ & $0.5 \pm 1.3$ \\
P081404803101 & 61098.86 & 6.52 & -- & -- & $-0.008 \pm 0.020$ & $-0.32 \pm 0.85$ & -- & -- \\
P081404803301 & 61100.64 & 4.94 & -- & -- & -- & -- & $-0.084 \pm 0.013$ & $-2.55 \pm 0.40$ \\
\end{longtable}
\normalsize
\end{center}

\end{document}